\documentclass[twocolumn,showpacs,preprintnumbers,amsmath,amssymb,superscriptaddress,balancelastpage,prb]{revtex4}

\include{options}
\usepackage {epsfig,color}
\usepackage{graphicx}

\begin{document}

\title{ Polaronic approach to strongly correlated electron system with strong electron-phonon interaction}

\author{I.~A.~Makarov}
\author{E.~I.~Shneyder}
\author{P.~A.~Kozlov}
\author{S.~G.~Ovchinnikov}
	\email{sgo@iph.krasn.ru}
\affiliation{Kirensky Institute of Physics SB RAS, 660036 Krasnoyarsk, Russia}

\date{\today}

%===========================================================================

\begin{abstract}

The three band $p-d$ model of strongly correlated electrons interacting with optical phonon via diagonal and off-diagonal electron-phonon interaction is considered within cluster perturbation theory. At first step the exact diagonalization of the Hamiltonian of CuO$_4$ cluster results in the construction of local polaronic eigenstates  $ \left| p \right\rangle $ with hole numbers $n_h=0,1,2$ per unit cell. The inter cluster hoppings and interactions are exactly written in terms of Hubbard operators $ X_f^{pq}= \left| p \right\rangle \left\langle q \right| $ determined within the multielectron polaronic eigenstates  $ \left| p \right\rangle $ at site $\bf{f}$. The Fermi type single electron quasiparticle dispersion and spectral weight are calculated for the undoped antiferromagnetic parent insulator like La$_2$CuO$_4$. The quasiparticle dispersion of Hubbard polarons is determined by a hybridization of the several Hubbard subbands with local Franck-Condon resonances. For small electron-phonon interaction the conductivity band is stronger renormalized then the valence band. Nevertheless for large electron-phonon interaction both bands are strongly renormalized with quasiparticle localization. Effect of partial compensation of diagonal and off-diagonal electron-phonon interaction at intermediate coupling is found.

\end{abstract}

%===========================================================================

% 71.38.-k	Polarons and electron-phonon interactions
% 74.25.Kc	Phonons (in 74.25.-q	Properties of superconductors)
% 74.72.-h Cuprate superconductors
% 71.10.Fd	Lattice fermion models (Hubbard model, etc.)
% 71.10.Hf	Non-Fermi-liquid ground states, electron phase diagrams and phase transitions in model systems

\maketitle

%===========================================================================
\section{Introduction \label{sec_introduction}}
%===========================================================================
The electronic structure of the undoped cuprates which are parent compounds for the high temperature superconductors is one of the most important problem for understanding the physics of the high-$T_c$ superconductivity~\cite{Damascelli2003}. Strong electron correlations (SEC) are known to result in the Mott insulating state. The polaron formation and strong electron-phonon interaction (EPI) are expected from the ionic nature of the undoped cuprates. The interplay between EPI and SEC is a key to resolve the quantum dynamics of the doped holes or electrons into the high-$T_c$ cuprates.

Experimental indications for strong EPI have been provided by angle-resolved photoemission spectroscopy (ARPES) studies of the undoped Ca$_2$CuO$_2$Cl$_2$ where broad Gaussian spectral features are found and interpreted in terms of Franck-Condon processes and polaron physics.~\cite{ShenKM2004,ShenKM2007} The large isotope effect on $T_c$ for underdoped cuprates and on the superfluid density at the optimal doping also suggests the importance of the EPI.~\cite{Hofer2000,Shneyder2010} In addition the ARPES data in doped cuprates have revealed the electron dispersion strongly renormalized by EPI.~\cite{Damascelli2003}

The interplay between charge, spin and lattice degrees of freedom has been studied theoretically by different approaches: the diagrammatic Monte Carlo method,~\cite{Mishchenko2004} the exact diagonalization within the limited functional space,~\cite{Bonca2008} the dynamical mean-field calculations within the Hubbard-Holstein model,~\cite{Sangiovanni06,Macridin2006,Rosch2004,Macridin2012} and the determinant quantum Monte Carlo method,~\cite{Johnston2013} see the review paper.~\cite{Mishchenko2009}

To consider both the local effects of the strong EPI as a set of the Franck-Condon resonances~\cite{Sawatzky1989} and the electron dispersion in the infinite lattice we present in this paper the polaronic version of the multielectron generalized tight-binding (GTB) approach to the quasiparticle band structure in SEC materials.
The generalized tight-binding method~\cite{Gav2000} and its \textit{ab-initio} version LDA+GTB~\cite{LDA+GTB} have been developed previously to study the electronic structure of the cuprates, cobaltites and manganites within the multiband Hubbard model. The polaronic version of the GTB method (p-GTB) as well as the initial GTB is an example of cluster perturbation theory.~\cite{Ovchinnikov89,Maier2005} The general idea of the GTB is the following: the initial multiband Hubbard-like Hamiltonian is written as the sum of the intracell part $H_c$ and the intercell hoppings and interactions $H_{cc}$. The exact diagonalization of the $H_c$ provides a set of local multielectron eigenstates  $ \left\{ {\left| p \right\rangle} \right\} $ and the intracell Hubbard operators $ X_f^{pq} = \left| p \right\rangle \left\langle q \right| $.~\cite{Hubbard65} Hence for, all two-site intercluster contributions in $H_{cc}$ will be given by a bi-linear products of Hubbard operators of different unit cells. In other words, any multiband Hubbard like model in GTB looks like the original Hubbard model. The only difference is the size of the matrix ${\left| p \right\rangle \left\langle q \right|}$. In the original Hubbard model with $4$ local states it is the $4 \times 4$ matrix. In the $5$-band $p-d$ model~\cite{Gav2000} with number of local electronic states $N_e \approx 100 $ the $\left| p \right\rangle \left\langle q \right|$ matrix size is $N_e \times N_e$. In polaronic approach~\cite{Mishchenko2004,Mishchenko2009} with number of coupled phonons $N_{ph}>10$ the size of ${\left| p \right\rangle \left\langle p \right|}$ matrix may be $1000 \times 1000$. Nevertheless for the low energy physics only a small part of multielectron states is relevant that is very important for the p-GTB approach below.

Besides the local states of Cu $d$-holes and O $p$-holes and local Coulomb interactions $U_{d}$, $U_{p}$, and $V_{pd}$ we include the electron-phonon interaction with optical phonon mode. Exact diagonalization of the Hamiltonian $H_c$ results in the local polaronic states $ \left\{ {\left| p \right\rangle} \right\} $ defined for different number of holes per unit cell $n_h=0,1,2$. For the three band $p-d$ model~\cite{Varma1987,Emery1987} with EPI the same procedure has been carried out in the Ref.~[\onlinecite{Piekarz1999}]. Within the GTB perturbation treatment of the intercluster Hamiltonian $H_{cc}$ we have calculated the dispersion of Hubbard polarons that is formed by hybridization of the Hubbard subbands with the local Franck-Condon resonances. To distinguish standard polaron for non- or weakly correlated electrons from polaron in strongly correlated electronic systems we introduce the notion of Hubbard polarons for the latter case.  The first doped hole or electron are delocalized at the weak EPI and becomes localized at the strong EPI. Effects of the diagonal (Hubbard-Holstein model) and off-diagonal EPI have been discussed. Partial compensation of these two EPI has been found at the intermediate coupling strength.

The organization of this work is as follows. In the next section we present the three band $p-d$ model Hamiltonian with additional bare phonons and EPI and separate the total Hamiltonian to a sum of individual unit cell part $H_c$ and the intercell contribution $H_{cc}$. Section~\ref{sec_pGTB} clarifies briefly the general ideas of polaronic version of the GTB formalism. In section~\ref{exact_diag} we discuss the structure of local Hilbert space with multiphonon and multielectron eigenstates with relevant for cuprates quantum numbers of holes per cell $n_h=0,1,2$. Sections~\ref{band_str} and~\ref{sp_fun} contain respectively the calculated polaronic band structure and spectral function for undoped cuprates. Discussion of the results is presented in Section~\ref{discussion}.

%===========================================================================
\section{The minimal model\label{MinModel}}
%===========================================================================
Among the different groups of high-temperature superconductors the copper oxides can be classified as the most correlated ones.
To consider interplay between strong EPI and strong electron correlations in these compounds we start from the minimal semirealistic model which describes holes in the CuO-plane interacting with the longitudinal optical vibrational mode:
\begin{eqnarray}
\label{tot_Hamiltonian}
H=H_{pd}+H_{ph}+H_{epi}.
\end{eqnarray}
The Hamiltonian $H_{pd}$ of three-band $p-d$ model~\cite{Emery1987,Varma1987} includes only orbitals that are most essential for the low-energy physics of cuprates, i.e. bonding Cu-$d_{x^2-y^2}$ and O-$p_x, p_y$ ones:
\begin{widetext}
\begin{eqnarray}
\label{pd_Hamiltonian}
H_{pd} & = & \sum\limits_ {{\bf f}, \sigma}\left(\varepsilon_ {d}^{_{}}-\mu \right) d_{{\bf f} \sigma}^\dag d_{{\bf f} \sigma}^{_{}} + \sum\limits_ {\alpha, {\bf g}, \sigma}\left(\varepsilon_ {p}^{_{}}-\mu \right) p_{ \left( \alpha \right) {\bf g} \sigma}^\dag p_{ \left( \alpha \right) {\bf g} \sigma}^{_{}} +
\sum\limits_ {\alpha, \alpha', {\bf g} \neq {\bf g}', \sigma} t_ {p_{_{\alpha}} p_{\alpha' }}^{_{}} \left( p_ { \left( \alpha \right) {\bf g} \sigma}^\dag p_ { \left( \alpha' \right) {\bf g}' \sigma}^{_{}} + h.c. \right) + \nonumber \\
& + & \sum\limits_ {\alpha, {\bf f} \neq {\bf g}, \sigma} t_ {p_{_{\alpha} } d}^{_{}} \left( d_ {{\bf f} \sigma}^\dag p_ { \left( \alpha \right) {\bf g} \sigma}^{_{}} + h.c. \right) +
\frac {1}{2} \sum\limits_ {{\bf f}, \sigma}  U_ {d}^{_{}} n_ { \left( d \right) {\bf f}}^{\sigma} n_ {\left( d \right) {\bf f} }^{-\sigma} + \frac {1}{2} \sum\limits_ {{\bf g}, \sigma}  U_ {p}^{_{}} n_ { \left( p_{ \alpha } \right) {\bf g} }^{\sigma} n_ {\left( p_{\alpha} \right) {\bf g} }^{-\sigma} +
\sum\limits_ {\alpha, {\bf f},{\bf g}, \sigma, {\sigma}'} V_ {p d}^{_{}} n_{\left( d \right) {\bf f} }^{\sigma} n_{ \left( p_{\alpha} \right) {\bf g} }^{{\sigma}'}.
\end{eqnarray}
\end{widetext}
Here indexes ${\bf f}$ and ${\bf g}$ run through positions of copper and oxygen plane atomic orbitals so that ${\bf g} = {\bf f} + {\bf r}_l^{_{}} $ and index $l$ enumerates the oxygen atoms in the tetragonal unit cell centered on site ${\bf f}$. Values $ \varepsilon_ {d} $ and $ \varepsilon_ {p} $ are the copper and oxygen hole energy levels respectively, and $\mu$ is the chemical potential. Operators $d_{{\bf f}, \sigma}$ and $p_{\left( \alpha \right){\bf g}, \sigma}$ describe the destruction of the holes with spin $\sigma$ at orbitals $d_{x^2-y^2}$ and $p_{\left( \alpha \right)} = \{ p_x,p_y \}$; $n_ {\left( d \right) {\bf f} }^{\sigma}$ and $n_ { p_{\left( \alpha \right)}  {\bf g}}^{\sigma}$ are the corresponding hole number operators. The hopping $t_ {p_{_{\alpha}} p_{\alpha' }}$ is between two nearest-neighbor oxygen sites, while $t_ {p_{\alpha} d}$ corresponds to the hopping between neighboring copper and oxygen orbitals. The explicit form  of hopping terms depends on the chosen phase condition~\cite{Jefferson92} and is consistent with the paper.~\cite{Gav2000}
The largest energy scale in the problem is defined by the set of the Coulomb repulsion parameters which consists of intra-atomic interactions $U_d$ and $U_p$ and the nearest-neighbor copper-oxygen Coulomb parameter $V_ {pd}$.

\begin{table} [b]
\begin{ruledtabular}
\begin{tabular}{c@{\hspace{3mm}}c@{\hspace{5mm}}*{12}{c@{\hspace{3mm}}}}
 Parameter &   &$\varepsilon_{d}$ & $\varepsilon_{p}$ &  $t_{pd}$ &  $t_{pp}$  \\[2mm]
\hline
value, eV  &   &             0    &   0.91            &  1.36     &    0.86    \\[2mm]
\end{tabular}
\caption{Hopping parameters and single-electron energies of the $p-d$ model~(\ref{pd_Hamiltonian}) obtained in the framework of the LDA+GTB method~\cite{LDA+GTB} and used in the present paper.}
\label{parameters}
\end{ruledtabular}
\end{table}

The \textit{ab initio} hopping parameters and single electron energies of the underlying $p-d$ model~(\ref{pd_Hamiltonian}) have been obtained~\cite{LDA+GTB} earlier for La$_2$CuO$_4$ using a Wannier function projection procedure~\cite{wf_Anisimov}. Their values are listed in Table~\ref{parameters}.
It should be noted that parameters in this Table differ from the values in Table III of Ref~[\onlinecite{LDA+GTB}] since we do not normalize them by $t_{pd}$. The Coulomb parameters are taken from the Ref.~[\onlinecite{wells1}] where the following values have been obtained by fitting to experimental ARPES data: $U_d=9$, $U_p=4$, and $V_{pd}=1.5$ (all values are given in eV).

The next term in Eq.~(\ref{tot_Hamiltonian}) is the phonon Hamiltonian $H_{ph}$ which includes the only in-plane full-breathing oxygen vibrations. It is sufficient for the purpose of the paper although more realistic approach should take at least several modes with the largest coupling constant into account. In cuprates these modes~\cite{Pintschovius05,Giustino2008} involve breathing Cu-O stretching motions of the planar and apical oxygen atoms and out-of-plane buckling motions of the planar oxygen atoms. Nevertheless to reasonably simplify the problem the phonon Hamiltonian $H_{ph}$ has been written down as
\begin{eqnarray}
\label{ph_Hamiltonian}
H_{ph} & = & \hbar\omega_{br} \sum\limits_ {{\bf f}} \left( e_{{\bf f} }^\dag e_{{\bf f}}^{_{}} + \frac{1}{2} \right).
\end{eqnarray}
Operators $e_{{\bf f}}^{_{}}$ and $ e_{{\bf f} }^\dag $ describe the destruction and creation, respectively, of the local phonon at site ${\bf f} $ with the frequency $\omega_{br}$. They are linearly related to the displacement operator via the canonical transformation:
\begin{eqnarray}
\label{ph_bose}
{\bf u}_ {{\bf f}l}^{_{}} =\sqrt{\frac {\hbar} {2M\omega_{br}} } {\bf n}_{l}^{_{}} \left( {e_{\bf f}^\dag + e_{\bf f}^{_{}}} \right),
\end{eqnarray}
that convert a quadratic form of the initial vibration Hamiltonian to a diagonal one. Here ${\bf n}_l^{_{}}$ is the unit polarization vector, which is assumed to be a constant independent of reciprocal space vector ${\bf q}$ as well as the phonon frequency $\omega_{br}$.

The phonon-induced renormalization of the electron energies is given by the Hamiltonian $H_{epi}$
\begin{eqnarray}
\label{epi_Hamiltonian}
H_{epi} & = & \sum\limits_ {{\bf f}, \sigma} \sum\limits_{l}{\left( -1 \right)}^{S_l^{_{}}} g_ {d}^{_{}} u_{{\bf f}l}^{_{}} d_ { {\bf f} \sigma}^\dag d_ {{\bf f} \sigma}^{_{}} + \nonumber \\
&+& \sum\limits_ {{\bf f} \neq {\bf g}, \sigma} {\left( -1 \right)}^{1+S_l^{_{}}} g_ {pd}^{_{}} u_{{\bf f}l}^{_{}} \left( d_ {{\bf f} \sigma}^\dag p_{ {\bf g} \sigma}^{_{}} + h.c. \right).
\end{eqnarray}
The first diagonal term results from modulation of copper on-site energy by the oxygen displacements from their equilibrium positions.
Since the value $ \varepsilon_ {d} $ increases with the Cu-O bond length increasing, the index $S_l^{_{}} = 0$ for displacements $\left( {\bf f} {\pm} r_l^{_{}}{\bf e}_{x \left( y \right)}^{_{}} {\pm} \delta r_l^{_{}} {\bf e}_{x \left( y \right)}^{_{}} \right)$ and $S_l^{_{}} = 1$ for $\left( {\bf f} {\pm} r_l^{_{}}{\bf e}_{x \left( y \right)}^{_{}} {\mp} \delta r_l^{_{}} {\bf e}_{x \left( y \right)}^{_{}} \right)$ where ${\bf e}_x$ and ${\bf e}_y$ are the unit vectors in the directions $x$ and $y$.
The second term in Eq.~(\ref{epi_Hamiltonian}) is off-diagonal one and describes the dependence of Cu-O hopping energy on the length of Cu-O bond.

It is convenient to define the dimensionless parameter of electron-phonon interaction which will be used throughout this work as a measure of the electron-phonon coupling strength,
\begin{eqnarray}
\label{epi_Const}
\lambda_{d \left( pd \right)}^{_{}} =  \frac {g_ {d \left(pd \right)}^2} { 2M \omega^2 W }.
\end{eqnarray}
For the chosen set of parameters the bandwidth $W$ of the occupied valence band in La$_2$CuO$_4$ is equal to $2.2$~eV. The phonon frequency is taken as $\omega_{br} = 90$~meV in accordance with the measured~\cite{Pintschovius05} value. In the present paper the parameters $\lambda_{d \left( pd \right)}^{_{}}$ vary between $0$ and $0.5$ values.

To proceed in the scheme of generalized tight-binding method we need to separate the total Hamiltonian~(\ref{tot_Hamiltonian}) into the intracell and intercell parts. The concomitant problem of non-orthogonality of all operators related to the oxygen sites of adjacent cells is solved explicitly via the canonical transformation~\cite{Shastry89} that introduces new operators in $k$-space. The detailed description of this procedure for the Hamiltonian of the $p-d$ model can be found elsewhere.~\cite{Raimondi96,Gav2000} The new oxygen hole operators are a linear combination of the Fourier transforms of the original $p_{\left( x \right) {\bf q} \sigma}$ and $p_{\left( y \right) {\bf q} \sigma}$ orbitals:
\begin{eqnarray}
\label{Shastry}
a_{{\bf q}\sigma}&=&-\frac{i}{\mu_{{\bf q}}}\left( s_x p_{\left( x \right) {\bf q} \sigma} + s_y p_{\left( y \right) {\bf q} \sigma} \right), \nonumber \\
b_{{\bf q}\sigma}&=& \frac{i}{\mu_{{\bf q}}}\left( s_x p_{\left( x \right) {\bf q} \sigma} - s_y p_{\left( y \right) {\bf q} \sigma} \right),
\end{eqnarray}
where ${s_{x\left( y \right)}} = \sin \left( {\frac{{{q_{x\left( y \right)}}a}}{2}} \right)$, $a$ is a lattice parameter, and $\mu_{\bf q} = \sqrt{ s_x^2 + s_y^2 } $. In the coordinate space, the group orbitals $a_{{\bf q}\sigma}$ and $b_{{\bf q}\sigma}$ are Wannier-like oxygen wave functions of $a_{1g}$ and $b_{1g}$ symmetry, respectively, which are centered at the copper site ${\bf f}$ and spread over several neighboring sites.

To diagonalize the phonon part of the Hamiltonian~(\ref{tot_Hamiltonian}) the procedure similar to the Shastry canonical transformation has been successfully applied in Ref.~[\onlinecite{PiekarzPhysC}]. Fourier transforms of the new phonon operators are given by equations
\begin{eqnarray}
\label{PhOper}
A_{{\bf q}}&=&-\frac{i}{\mu_{{\bf q}}}\left( s_x e_{\left( x \right) {\bf q}} + s_y e_{\left( y \right) {\bf q}} \right), \nonumber \\
B_{{\bf q}}&=&-\frac{i}{\mu_{{\bf q}}}\left( s_y e_{\left( x \right) {\bf q}} - s_x e_{\left( y \right) {\bf q}} \right).
\end{eqnarray}

After the orthogonalization~(\ref{Shastry},\ref{PhOper}) the total Hamiltonian can be written as a sum of the individual unit cell part $H_{c}$ and the intercell contribution $H_{cc}$
\begin{widetext}
\begin{subequations} \label{Hc_Hcc}
\begin{align}
\label{Hc_Hcc_tot}
&H = H_c + H_{cc}, \qquad H_c = \sum\limits_ {{\bf f}, \sigma} H_{{\bf f} \sigma}, \qquad H_{cc} = \sum\limits_ {{\bf f}, {\bf f'}, \sigma} H_{{\bf f} {\bf f'} \sigma}, \\
\label{Hc}
H_{{\bf f} \sigma} = & \sum\limits_ {\beta} \left(\varepsilon_ {\beta}-\mu \right) n_{\left( \beta \right) {\bf f} }^{\sigma} - 2{t}_{pd}^{_{}} \mu_{0}^{_{}}\left( {d_{{\bf f} \sigma}^\dag b_{{\bf f} \sigma}^{_{}}+ h.c.}  \right)
+\frac{1}{2} \sum\limits_ {\beta} {U}_{\beta}^{_{}} n_ { \left( \beta \right) {\bf f} }^{\sigma} n_ {\left( \beta \right) {\bf f} }^{-\sigma}
+ \sum\limits_ {\sigma'} \tilde{V}_{pd}^{_{}} n_ { \left( d \right) {\bf f} }^{\sigma} n_ {\left( b \right) {\bf f} }^{\sigma'} + \nonumber \\
& + \hbar\omega_{br} A_{{\bf f}}^\dag A_{{\bf f}}^{_{}} + 2{\lambda}_{d}^{_{}} \mu_{0}^{_{}} \left( { A_{{\bf f}}^\dag + A_{{\bf f}}^{_{}} }\right) n_ { \left( d \right) {\bf f} }^{\sigma} + 2{\lambda}_{pd}^{_{}} \rho_{0}^A \left( { A_{{\bf f}}^\dag} + A_{{\bf f}}^{_{}} \right) \left( d_{{\bf f} \sigma}^\dag b_{{\bf f} \sigma}^{_{}}+ h.c. \right),\\
\label{Hcc}
H_{{\bf f} {\bf f'} \sigma}  = & -2{t}_{pd}^{_{}} \mu_{\bf{ff'}}^{_{}} \left( d_{{\bf f} \sigma}^\dag b_{{\bf f'} \sigma}^{_{}}+ h.c. \right)-2t_{pp}^{_{}} \nu_{\bf{ff'}}^{_{}} \left( b_{{\bf f} \sigma}^\dag b_{{\bf f'} \sigma}^{_{}}+ h.c. \right) + \nonumber \\
& + 2{\lambda}_{d}^{_{}}  \mu_{\bf{ff'}}^{_{}} \left( { A_{{\bf f}}^\dag} + A_{{\bf f}}^{_{}} \right) n_ { \left( d \right) {\bf f'} }^{\sigma}  + 2\lambda_{pd}^{_{}} \rho_{\bf{ff'h}}^A \left( { A_{{\bf f}}^\dag} + A_{{\bf f}}^{_{}} \right) \left( d_{{\bf f'} \sigma}^\dag b_{{\bf h} \sigma}^{_{}}+ h.c. \right).
\end{align}
\end{subequations}
\end{widetext}
Here index $\beta = \{d,b\}$ enumerates the plane orbitals of copper and oxygen, respectively. The terms containing operators of $a_{1g}$ oxygen orbital as well as $B$-type phonon excitations are omitted in the Hamiltonian~({\ref{Hc_Hcc}}) because of their insignificantly small influence on the low lying local eigenstates.
Expressions for the new parameters and renormalizing coefficients from Eq.~({\ref{Hc_Hcc}}) are presented in the Appendix~A.

We should to emphasize that electron-phonon interaction in the derived Hamiltonian has the non-local character. The procedure of orthogonalization results in renormalization of all matrix elements in Eq.~({\ref{Hc_Hcc}}) which become strongly distance-dependent even for the initial parameters are taken to be non zero only for the nearest neighbors.

%===========================================================================
\section{The polaronic version of the GTB method \label{sec_pGTB}}
%===========================================================================
The GTB method has been proposed to calculate the band structure of compounds in the limit of strong electron correlations. When Coulomb energy is much higher then kinetic energy the idea of bare electron as zero approximation of a theory does not work. Contrary to the conventional tight-binding method the local states in GTB approach are not free electron ones but rather quasiparticle excitations between multielectron terms of $d^n$ and $d^{n\pm1}$ configurations.

It should be clarified what these quasiparticles are. The localized multielectron $d^n$ configurations for a separate ion in the crystal field can be easily determined from local electroneutrality. Lets denote $m$ possible for the given configuration $d^n$  terms as $E_m\left(n\right)$. One of them is the ground term $E_0\left(n\right)$ that is occupied at zero temperature. An excitations from the ground to any term with the energy $\Delta E_{m0}=E_m\left(n\right) - E_0\left(n\right)$ are the local Bose-type quasiparticles such as excitons, magnons and so on.
If external electron comes to the given ion the later changes its configuration to $d^{n+1}$ with its own energy spectra $E_{m'}\left( n + 1 \right)$. The electron addition energy $\Omega_{mm'}=E_m \left(n+1 \right) - E_{m'}\left(n\right)$ may be considered as the single particle excitation between two multielectron configurations, with the initial state $E_m\left(n\right)$ and the final state $E_{m'}\left(n+1\right)$.
The interatomic interactions transform these local excitation energies $\Omega_{mm'}$ into the quasiparticle bands $\Omega_{mm'} \left( \bf{q} \right)$.

Due to the large number of initial and final states there are different quasiparticles with all possible pairs of $\left( m,m' \right)$.
It is evident that contribution of each particular $\left| m,n \right\rangle \to \left| m',n+1 \right\rangle$ excitation is determined by the corresponding matrix element $ \left\langle m,n+1 \right| c^\dag  \left| m',n \right\rangle $ of the electron creation operator $ c^\dag $. Moreover excitations from empty $\left| m,n \right\rangle$ to empty $\left| m',n \right\rangle$ states have zero spectral weights while their energies $\Omega_{mm'}$ are defined. Non zero spectral weight of the quasiparticles results from total or partial occupation of participating multielectron terms.

Therefore in GTB picture the correlated electron is treated as a linear combination of different quasiparticles and more important that each of them has its own quasiparticle weight. It results in the crucial difference between free electron picture and GTB one. The following spectral weight redistribution over these quasiparticles defines the underlying effects of the band structure formation in correlated systems.

The polaronic version of the GTB method is the natural development of the approach to the systems with strong electron-electron and electron-phonon interactions.
While in GTB picture the quasiparticles are formed via the electron-electron interactions in p-GTB method the formation of quasiparticles is owing to the presence of both the electron-electron and electron-phonon interactions.

%===========================================================================
\section{Exact multielectron and multiphonon eigenstates of $\bf{CuO_4}$ cluster \label{exact_diag}}
%===========================================================================
Undoped La$_2$CuO$_4$ has a mixture of $d^9p^6$ (hole on copper) and $d^{10}p^5$ (hole on oxygen) configurations. Both are related with one hole per unit cell (here CuO$_4$ cluster). The electron addition results in the $d^{10}p^6$ local configuration without holes. The electron removal results in a mixture of two-hole local configurations $d^9p^5$,  $d^{10}p^4$,  and $d^8p^6$. That is why one has to start the GTB procedure with exact diagonalization of the local Hamiltonian $H_c$ given by~(\ref{Hc_Hcc}) in the $3$ subspaces of the Hilbert space with the number of holes ${n_h} = 0,1,2$.

%===========================================================================
\subsection{Subspace with $\bf{{n_h} = 0}$ \label{nh0}}
%===========================================================================
The eigenstates can be written as
\begin{equation}
|0,\nu  \rangle  =  |0\rangle |\nu \rangle, \nu  = 0,1,...,N_{max}.
\label{es0}
\end{equation}
Here $|0\rangle $ means the hole vacuum, corresponding to the electronic configuration $\left| {{d^{10}}{p^6}} \right\rangle $, $|\nu \rangle $ denotes muliphonon state with number of phonons ${n_{ph}} = \nu $ that results from $\nu$ $\times$ action of phonon creation operator ${A^ + }$ on vacuum state $|0\rangle $ of harmonic oscillator:
\begin{eqnarray}
\left| \nu  \right\rangle  = \frac{1}{{\sqrt {\nu !} }}{\left( {{A^ + }} \right)^\nu }\left| 0 \right\rangle.
\label{hos}
\end{eqnarray}

%===========================================================================
\subsection{Subspace with $\bf {n_h} = 1$ \label{nh1}}
%===========================================================================
\begin{figure*}
\center
\includegraphics[width=0.45\linewidth]{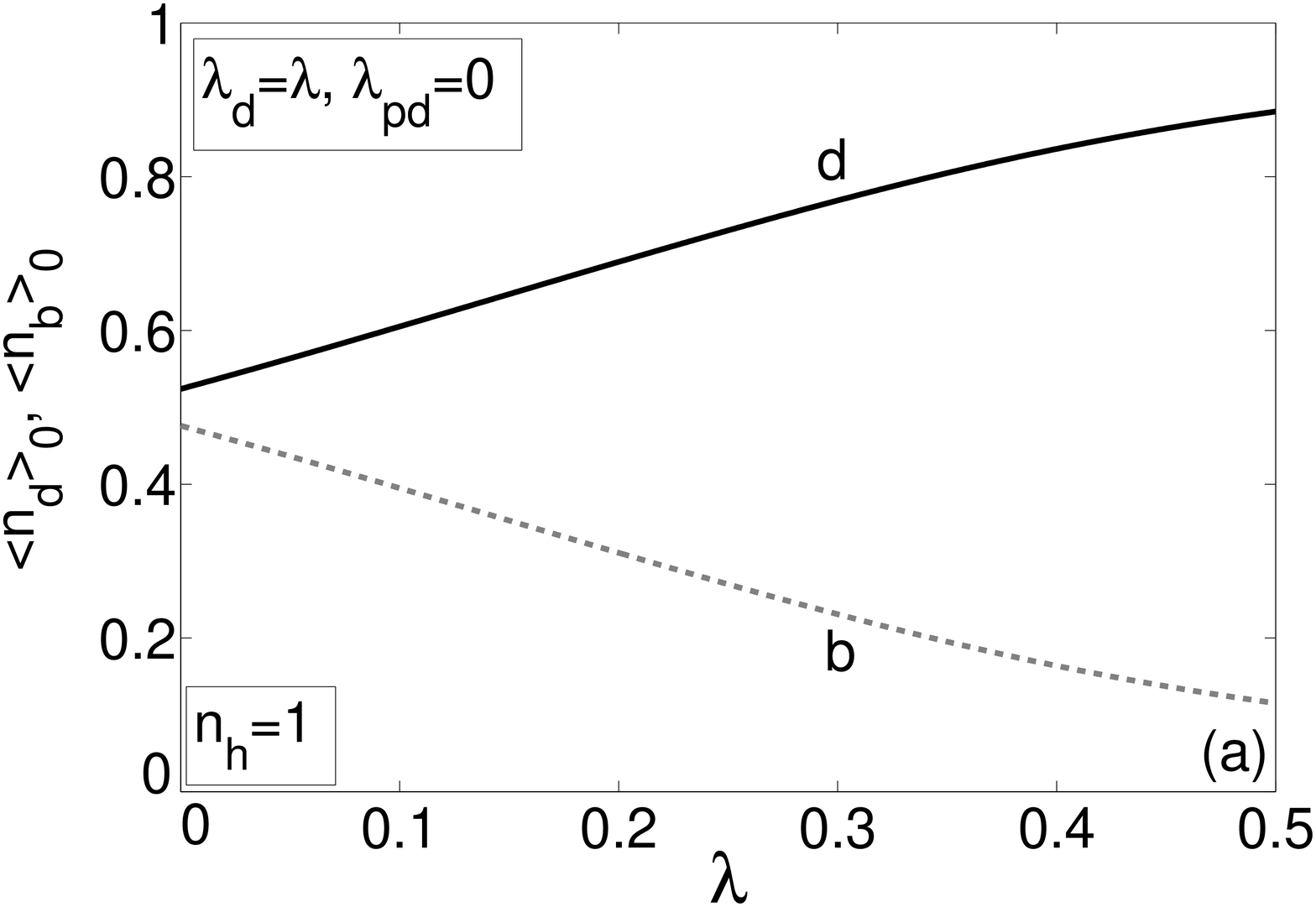}
\includegraphics[width=0.45\linewidth]{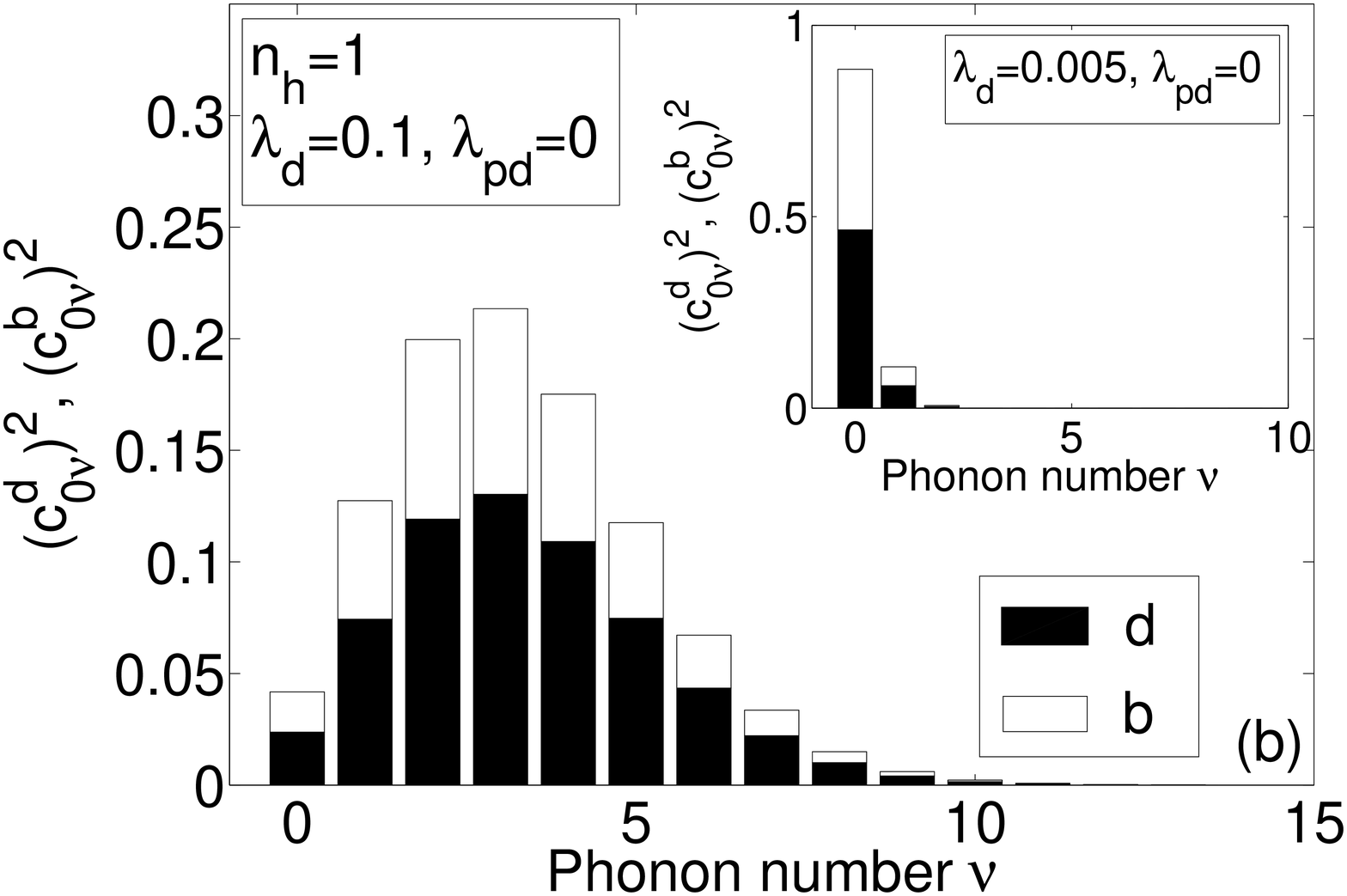}
\includegraphics[width=0.45\linewidth]{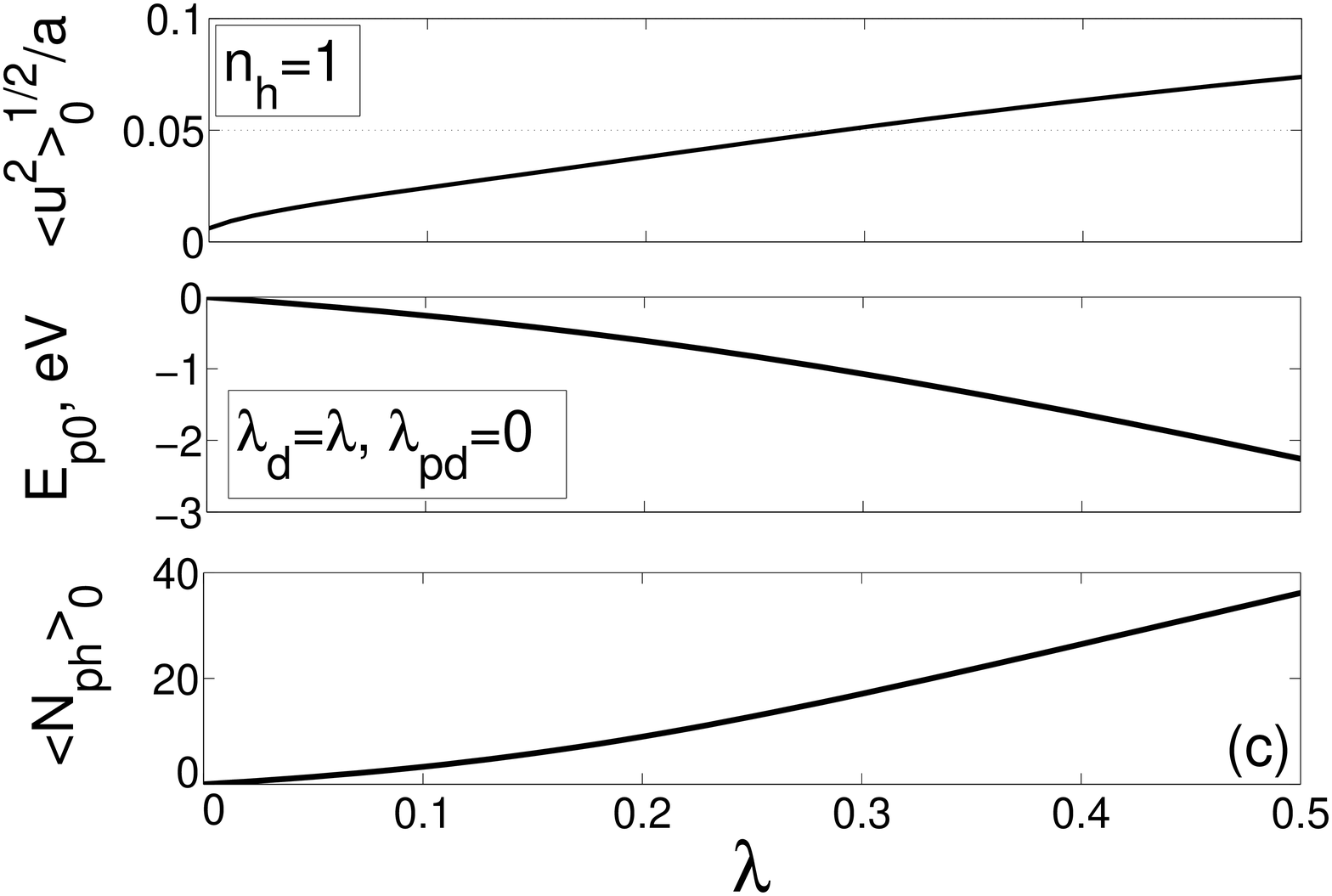}
\includegraphics[width=0.45\linewidth]{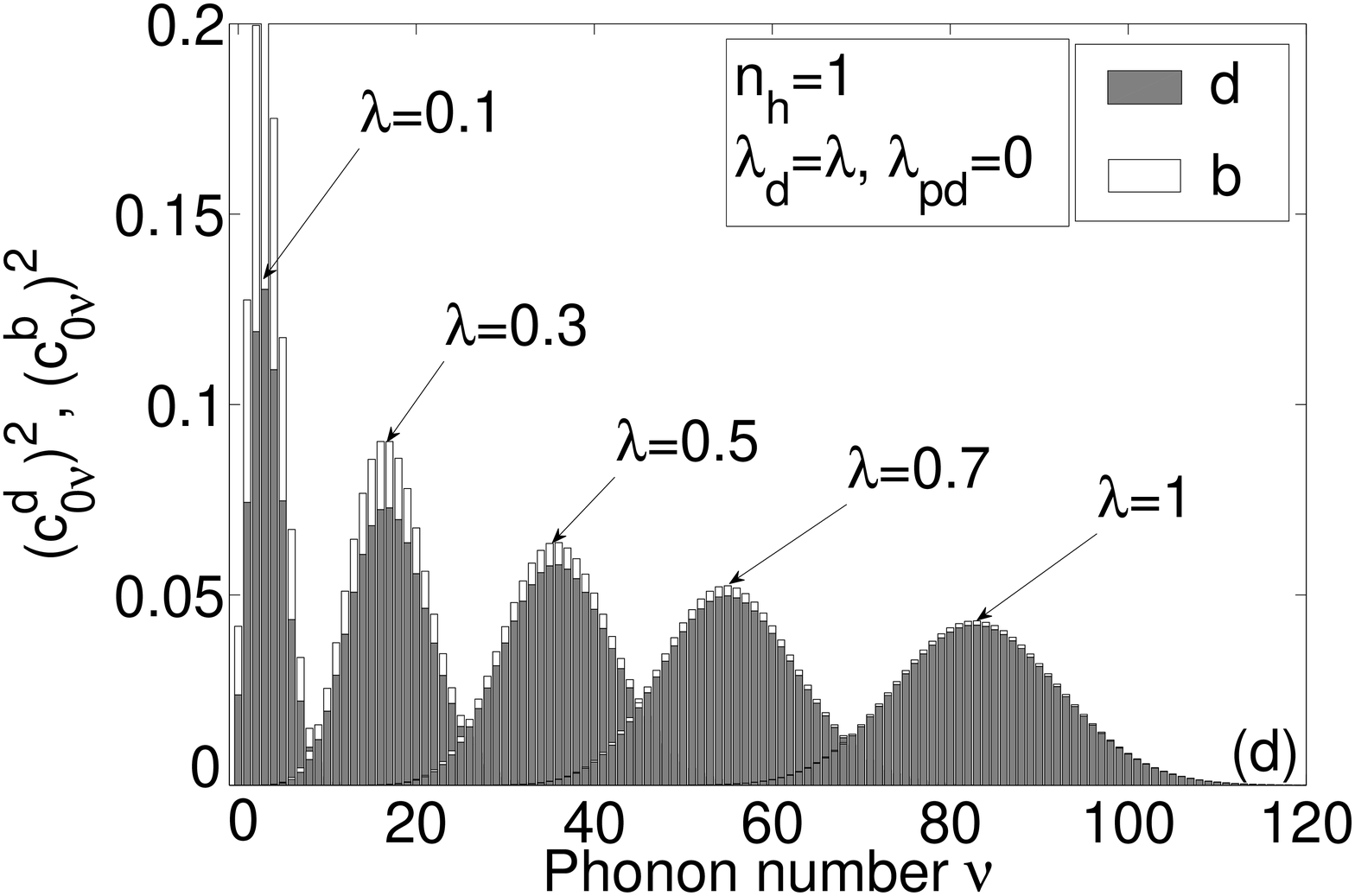}
\caption{\label{fig:fnh1} The distribution of a hole among copper $d$ and oxygen $b$ orbitals as function of (a) the diagonal EPI $\lambda_d$ and (b) the phonon number $\nu$  for small coupling and (d) the same for large coupling, (c) the ground state $\left| {1\sigma ,0} \right\rangle $ ratio of the square root of the average square of oxygen displacement to the lattice parameter ${\frac{\sqrt {{{\left\langle {{u^2}} \right\rangle }_0}} }{a}}$, polaronic shift $E_{p0}$, and number of phonons ${\left\langle {{N_{ph}}} \right\rangle _0}$ dependencies on the diagonal EPI.}
\end{figure*}
\begin{figure*}
\center
\includegraphics[width=0.45\linewidth]{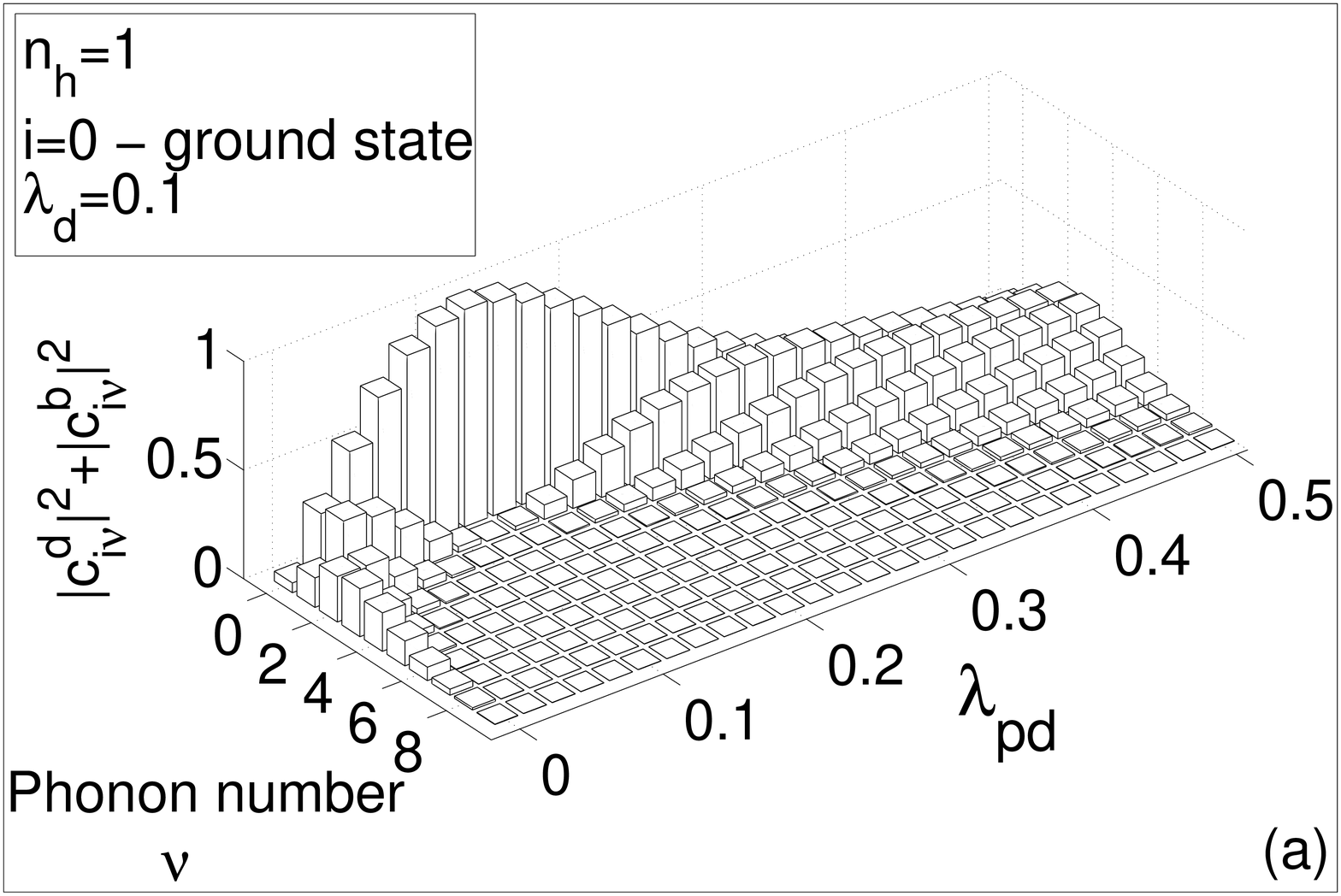}
\includegraphics[width=0.45\linewidth]{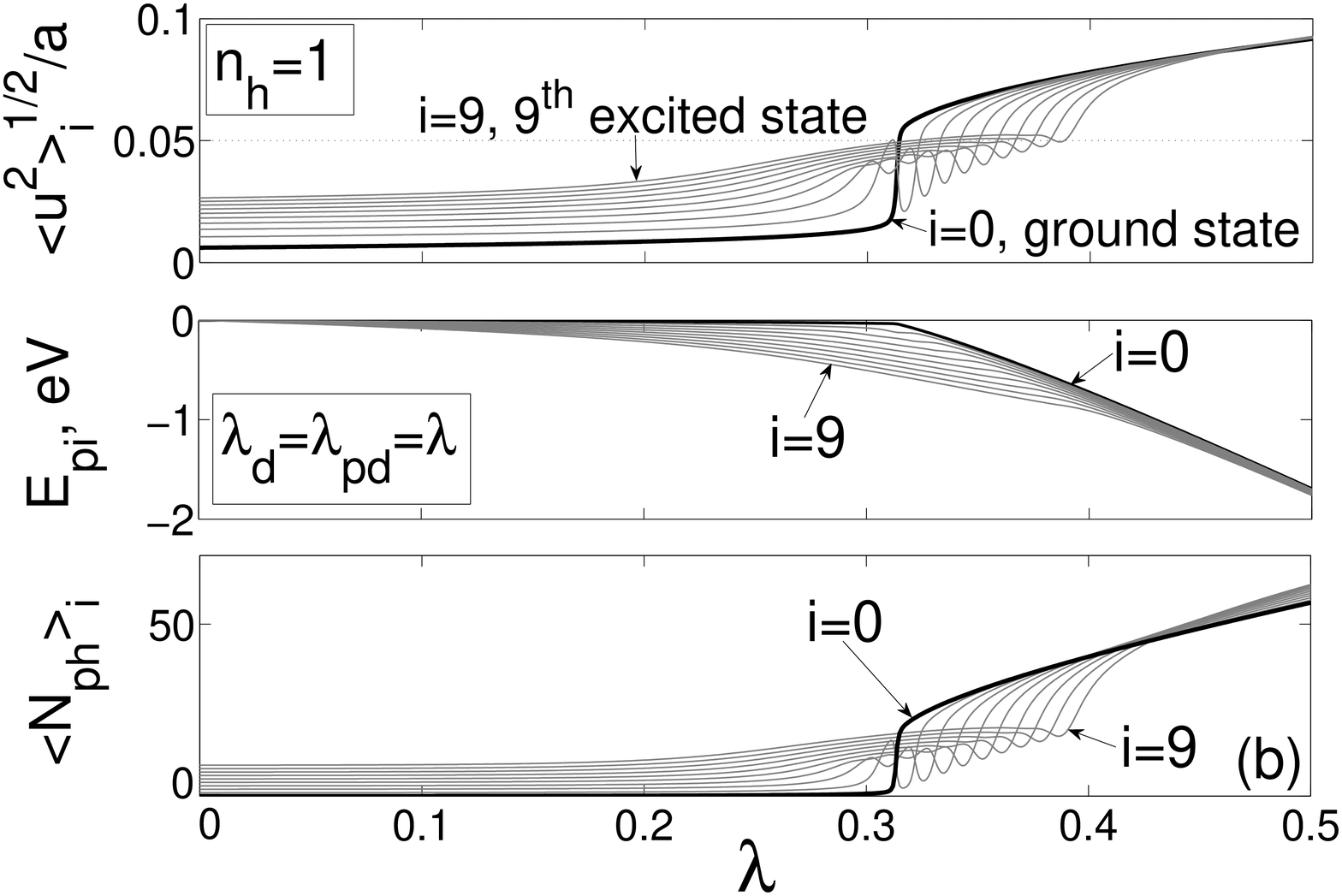}
\includegraphics[width=0.45\linewidth]{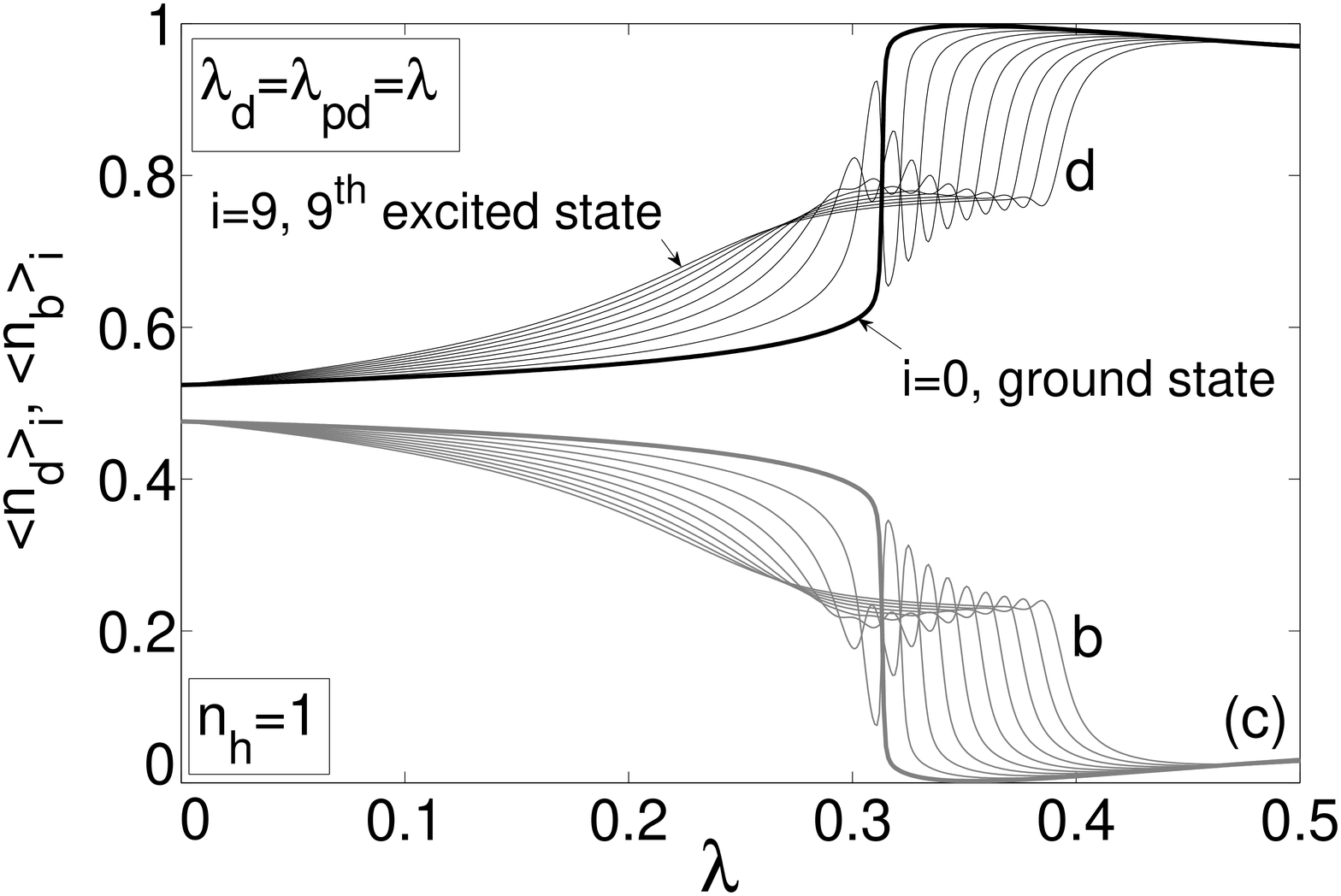}
\includegraphics[width=0.45\linewidth]{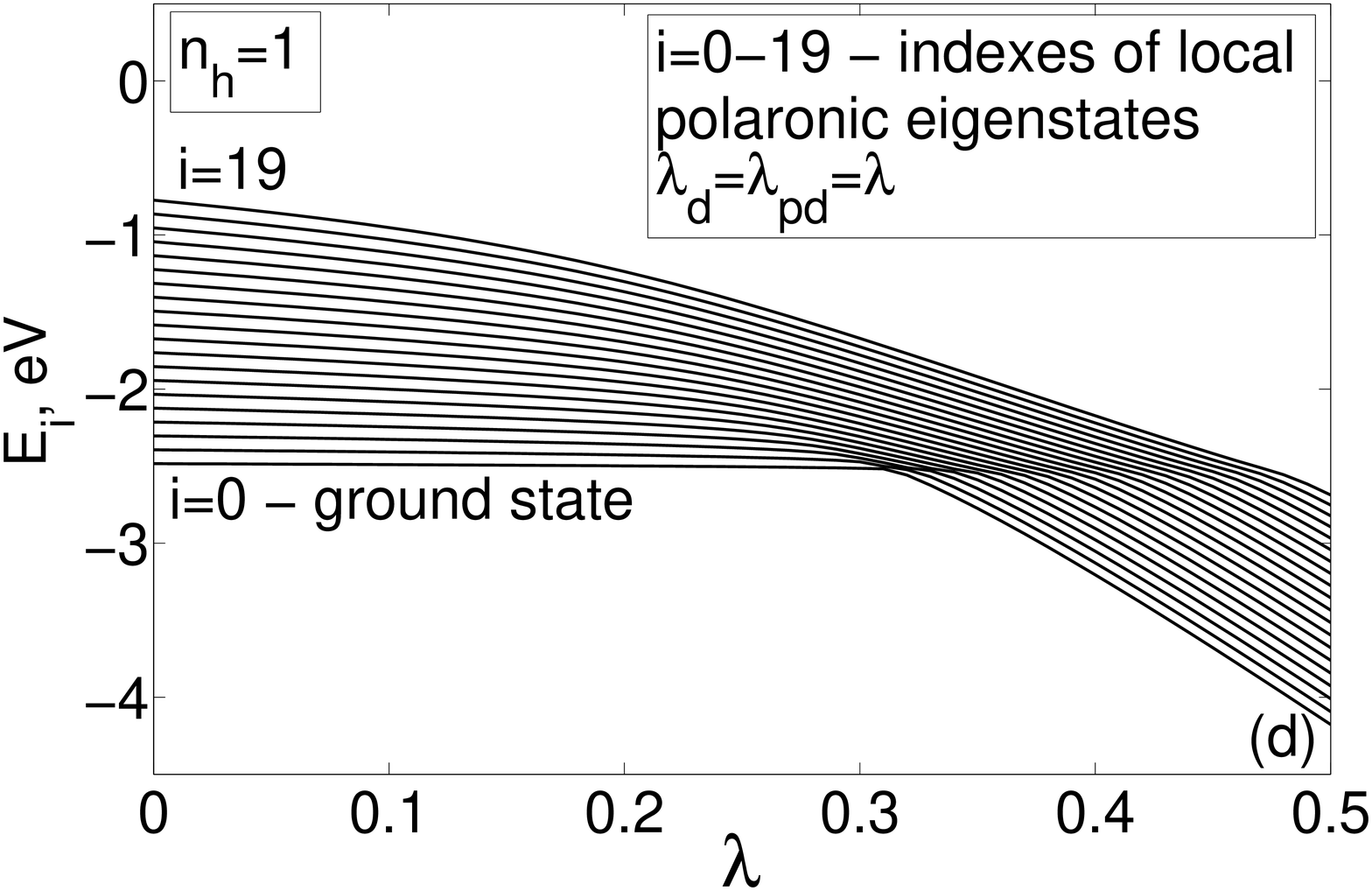}
\caption{\label{fig:fnh1_lpd} The effect of both diagonal and off-diagonal EPI on (a) the distribution of a hole as function phonon number and $\lambda_{pd}$, (b) ratio of the square root of the average square of oxygen displacement to the lattice parameter ${\frac{\sqrt {{{\left\langle {{u^2}} \right\rangle }_i}} }{a}}$, polaronic shift $E_{pi}$, and average number of phonons ${\left\langle {{N_{ph}}} \right\rangle _i}$, (c) the energy levels of a single hole ground and excited states, (d) the change of the copper and oxygen orbital occupations for the ground and excited states.}
\end{figure*}
\begin{figure}{R}
\includegraphics[width=1\linewidth]{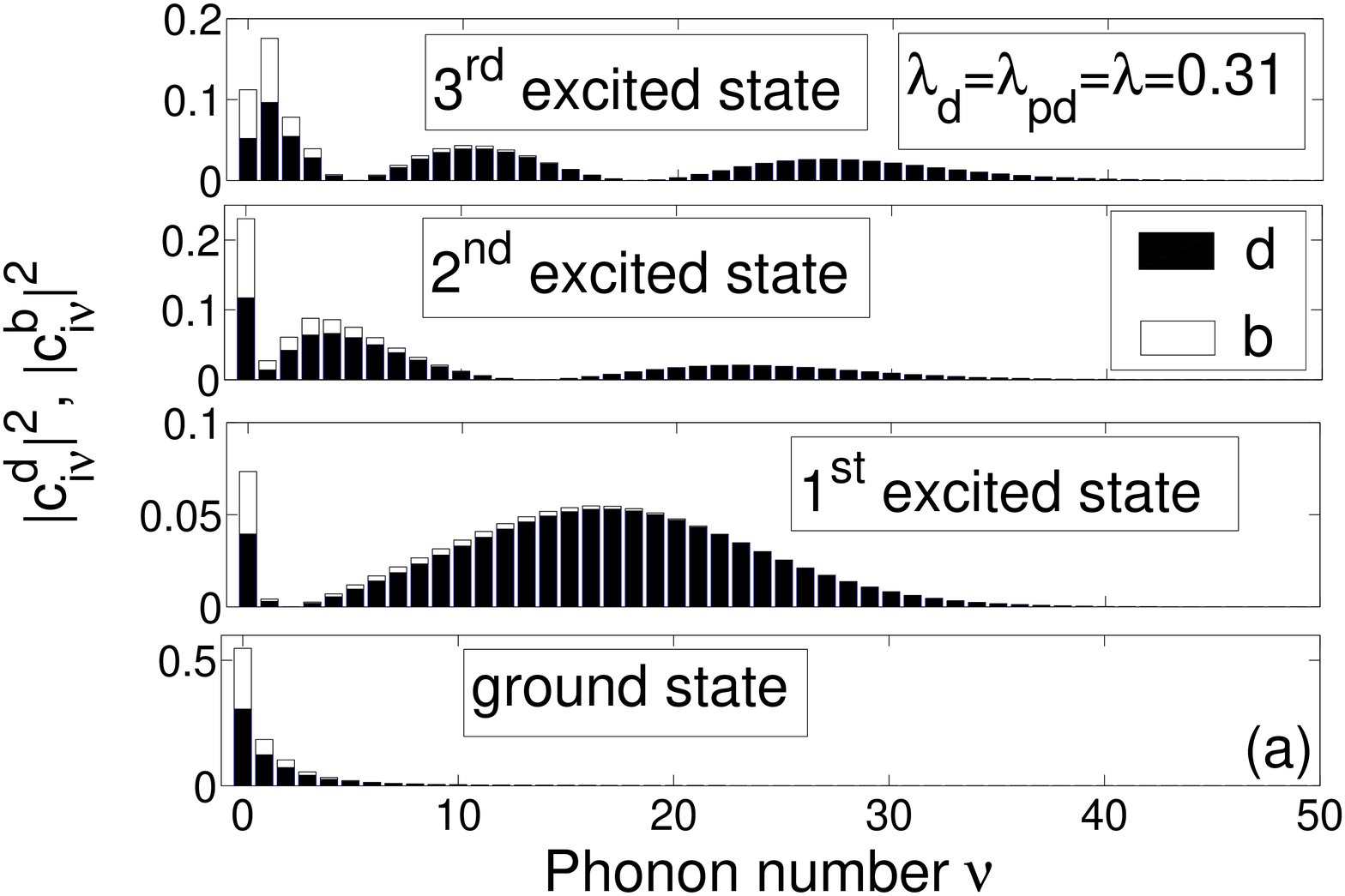}
\includegraphics[width=1\linewidth]{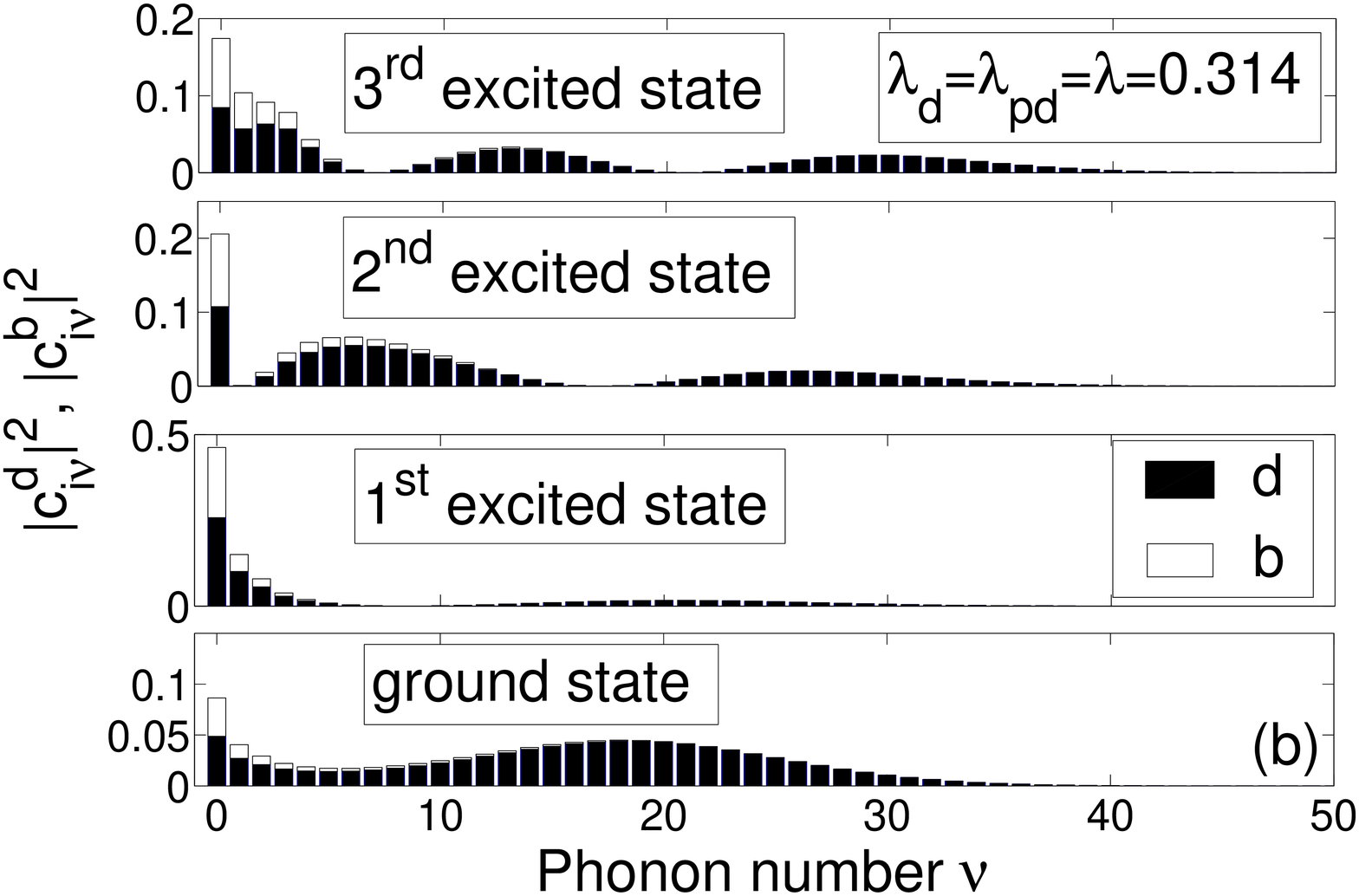}
\includegraphics[width=1\linewidth]{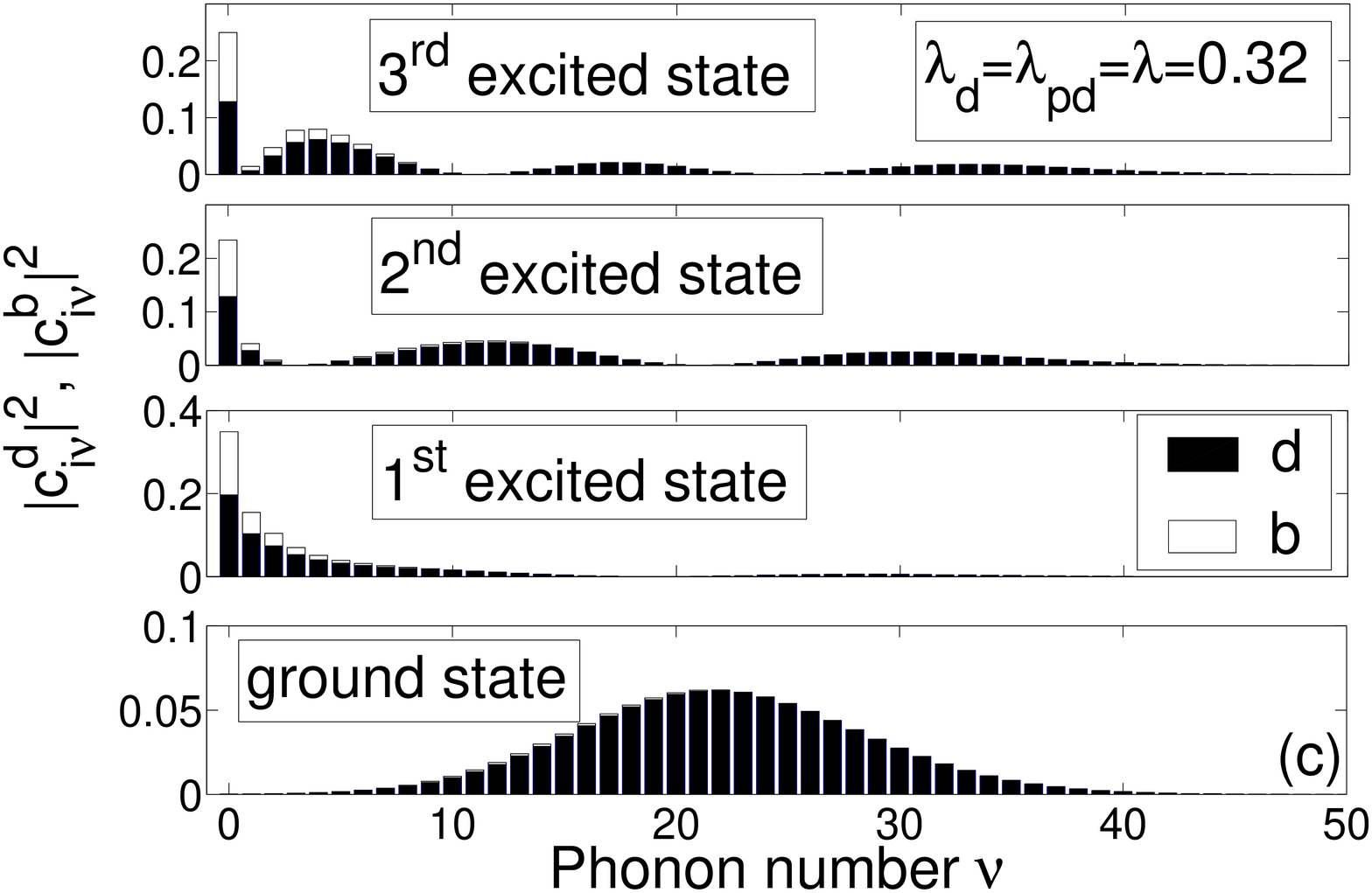}
\caption{\label{fig:crossover} The copper and oxygen hole distribution vs phonon number in the single-hole ground and three excited states for equal parameters of the diagonal and off-diagonal EPI (a) below  the critical value, (b) in the critical point and (c) above it.}
\end{figure}
Without EPI there are two eigenstates corresponding to the bonding and antibonding mixtures of $d_{x^2-y^2}$ copper and ${p_{x,y}}$ oxygen orbitals. The $i$-th cluster eigenstate with one hole is a spin doublet with projection of spin $\sigma$, that may be written in the following way
\begin{equation}
\left| {1\sigma ,i} \right\rangle  = \sum\limits_{\nu  = 0}^{{N_{\max }}} {\left( {c_{i\nu }^d\left| {{d_\sigma }} \right\rangle \left| \nu  \right\rangle  + c_{iv}^b\left| {{b_\sigma }} \right\rangle \left| \nu  \right\rangle } \right)}.
\label{es1}
\end{equation}
Here $\left| {{d_\sigma }} \right\rangle  = d_{\sigma }^ + \left| 0 \right\rangle $, $\left| {{b_\sigma }} \right\rangle  = b_\sigma ^ + \left| 0 \right\rangle $. In general, the ground and excited eigenstates~(\ref{es1}) characterize the electron surrounded by a cloud of phonons, e.g. the polaron. The polaronic shift $E_{pi}$ of the term~(\ref{es1}) may be calculated as a difference in the term $i$ energy with and without EPI: $E_{pi}\left( \lambda  \right) = {E_i}\left( \lambda  \right) - {E_i}\left( {\lambda  = 0} \right)$.~\cite{Lang_Firsov} It is instructive also to calculate a square root of the average square of oxygen displacement
\[\sqrt{{\left\langle {u_{{\bf{g}} l}^2} \right\rangle}_i}  = \sqrt{\left\langle {{{\left( {{n_h},i} \right)}_{\bf{g}l}}} \right|u_{{\bf{g}}l }^2\left| {{{\left( {{n_h},i} \right)}_{\bf{g}l}}} \right\rangle} \]
for the cluster centered at site $\bf{f}$ in the eigenstate ${\left| {{n_h},i} \right\rangle }$. This value allows to restrict the EPI parameters not to get the lattice melting following the Lindeman criteria. For values $\lambda_d,\lambda_{pd}<0.5$ the Lindeman criteria  $\sqrt{{\left\langle {u^2} \right\rangle}_0} \ll a$ is fullfilled~(Fig.~\ref{fig:fnh1}c).

For simplicity we start our discussion with the effect of diagonal EPI when the off-diagonal term $\lambda _{pd} = 0$. The monotonic distribution of one hole among copper and oxygen orbitals as function of  $\lambda_d$ is shown in~(Fig.~\ref{fig:fnh1}a). A structure of polaronic state~(\ref{es1}) is determined by the coefficients $c_{i\nu }^d$ and $c_{i\nu}^b$ that are shown in~(Fig.~\ref{fig:fnh1}b) for different phonon numbers. Without EPI there is only one summand in Eq.~(\ref{es1}) with $\nu  = 0$. For small EPI ${\lambda _d} = 0.01$ the maximal probability is for $0$-phonon state $\left( {{{\left| {c_{00}^d} \right|}^2} + {{\left| {c_{00}^b} \right|}^2}} \right)\left| 0 \right\rangle$, phonon cloud around electron is very thin.  For ${\lambda _d} = 0.1$ the maximal probability to find a hole either on copper or on oxygen occurs for three-phonon $\nu  = 3$ state with rather large contributions from $1$-, $2$- and $4$-phonon states, similar demonstration of polaronic effect on the local single-hole ground state has been obtained in the paper.~\cite{Piekarz1999} With increasing EPI the maximal probability shifts to the multiphonon states with $\nu  = 15$ for ${\lambda _d} = 0.3$ and $\nu  = 35$ for ${\lambda _d} = 0.5$~(Fig.~\ref{fig:fnh1}d).

The maximal number of phonons $N_{max}$ in~(Fig.~\ref{fig:fnh1}a) is a parameter of the theory however the choice of this parameter depends on value of $\lambda_{d\left(pd\right)}$. For any given EPI coupling we have calculated the hole distribution vs phonon number $\nu$ for different $N_{max}$, and increase the $N_{max}$ value up to stable zero contributions for higher phonon numbers. Thus, for ${\lambda _d} = 0.5$ $N_{max}=60$, and for ${\lambda _d} = 1$ $N_{max}=120$. For small ${\lambda _d}$ effect of EPI on the occupation number of $d$-orbital is weak because the polaronic shift $E_p$ is small vs covalence that is determined by $t_{pd}$, $\left| {{E_p}}\right| \ll {t_{pd}}$. Hole hopping from oxygen to copper  takes time $\approx{1 \mathord{\left/ {\vphantom {1 {{t_{pd}}}}} \right. \kern-\nulldelimiterspace} {{t_{pd}}}}$ and it is faster than polaron forms a potential well and localized in it for the time $\approx{1 \mathord{\left/ {\vphantom {1 {{E_p}}}} \right. \kern-\nulldelimiterspace} {{E_p}}}$. A charge carrier is untrapped by the deformation field and delocalized, we can call such state a large radii local polaron (large local polaron). Term  ``local" here indicates the state without dispersion. The polaronic dispersion will be considered in the next section, nevertheless classification of polarons as large and small radii corresponds to delocalized/localized states. The stronger EPI the larger is population of hole on copper orbital that results from the hole self-trapping, because oxygen holes are more mobile than copper ones. Smoothly with ${\lambda _d}$ growth at ${\lambda _d} = 0.03 - 0.04$ the phonon cloud transforms from the narrow distribution with the maximum at $0$-phonon component in the eigenstates~(\ref{es1}) to the multiphonon components. Nevertheless the partial occupation of oxygen orbital takes place for large EPI also. Only in the non-realistic limit ${\lambda _d} = 1$ we have found full occupation of the $d$-orbital~(Fig.~\ref{fig:fnh1}d).

Now we consider the effect of both diagonal ${\lambda_d}$ and off-diagonal $\lambda_{pd}$ EPI.
Contrary to the diagonal EPI the off-diagonal one has a trend for hole delocalization. It results in a partial compensation of both EPI contributions. For the same value of $\lambda$ the multiphonon contribution to the eigenstate~(\ref{es1}) is much smaller (compare Fig.~\ref{fig:fnh1_lpd}a and Fig.~\ref{fig:fnh1}b). From Fig.~\ref{fig:fnh1_lpd}a it is clear that increasing of ${\lambda _{pd}}$ up to the ${\lambda _d}$ value results in decreasing the size of the phonon cloud and the multiphonon  weight.  With parameters $\lambda_d,\lambda_{pd}$ there are too many variants, for simplicity we restrict ourselves further by the case ${\lambda_d} = {\lambda _{pd}} = \lambda $. In this case we have found a critical value ${\lambda_c} = 0.314$ separating properties of our local polaron.  At $\lambda  \ll {\lambda _c}$ the square root of average square of displacement $\sqrt{{\left\langle {u^2} \right\rangle}}$, number of phonons $\left\langle {{N_{ph}}} \right\rangle $, and polaronic shift $E_p$  are less than for the diagonal EPI regime (compare Fig.~\ref{fig:fnh1_lpd}b and Fig.~\ref{fig:fnh1}c).
This is effect of partial compensation of diagonal and off-diagonal EPI.
At small $\lambda$ the redistribution of hole between copper and oxygen vs $\lambda$ is almost absent (Fig.~\ref{fig:fnh1_lpd}c), polaronic shift of the ground and excited eigenstates~(\ref{es1}) is almost the same (Fig.~\ref{fig:fnh1_lpd}b), the number of phonons $\left\langle {{N_{ph}}} \right\rangle $ is close to the case of EPI absence, the difference $\left\langle {1,i + 1} \right|{N_{ph}}\left| {1,i + 1} \right\rangle  - \left\langle {1,i} \right|{N_{ph}}\left| {1,i} \right\rangle $ is close to $1$. These are the large local polaron states. Approaching the critical value ${\lambda _c}$ all characteristics of the ground and excited terms~(\ref{es1}) have drastically changed. Non monotonic dependence appears for the hole occupation numbers (Fig.~\ref{fig:fnh1_lpd}c), the average square of displacement ${\left\langle {u^2} \right\rangle}_i$, number of phonons ${\left\langle {{N_{ph}}} \right\rangle}_i$ (Fig.~\ref{fig:fnh1_lpd}b) in the ground and excited terms, the energies of the ground and excited terms are close to crossover (Fig.~\ref{fig:fnh1_lpd}d).
	
We have compared the polaronic effect in the ground and several excited states in Fig.~\ref{fig:crossover}. Below the critical value there is maximal contribution of $\nu  = 0$ in the ground state while the multiphonon contributions dominate in the excited states (Fig.~\ref{fig:crossover}a). At the critical value $\lambda  = {\lambda _c} = 0.314$ both $0$-phonon and multiphonon contributions exist in the ground state, one is typical for the large and the other for the small polaron(Fig.~\ref{fig:crossover}b). Simultaneously in the first excited state (Fig.~\ref{fig:crossover}b) we have noticed the increase in the $0$-phonon contribution vs the same excited state in Fig.~\ref{fig:crossover}a.

All these results indicate that at $\lambda  = {\lambda _c} = 0.314$ there is a crossover between large and small local polaron states. This critical behavior is different to the case of the diagonal EPI where the transformation between large and small local polaron states is smooth. For $\lambda  = 0.32$ in the ground state with ${n_h} = 1$ we have found almost full occupation of the copper orbital and empty oxygen one (Fig.~\ref{fig:fnh1_lpd}c,~\ref{fig:crossover}c). Further increase of $\lambda$ results in a small decreasing of copper hole occupation while the average square of displacement ${\left\langle {u^2} \right\rangle}_0$, number of phonons ${\left\langle {{N_{ph}}} \right\rangle}_0$, and polaronic shift $\left| {{E_{p0}}} \right|$ continue to grow.

%===========================================================================
\subsection{Subspace with $\bf {n_h} = 2$ \label{nh2}}
%===========================================================================
\begin{figure*}
\center
\includegraphics[width=0.45\linewidth]{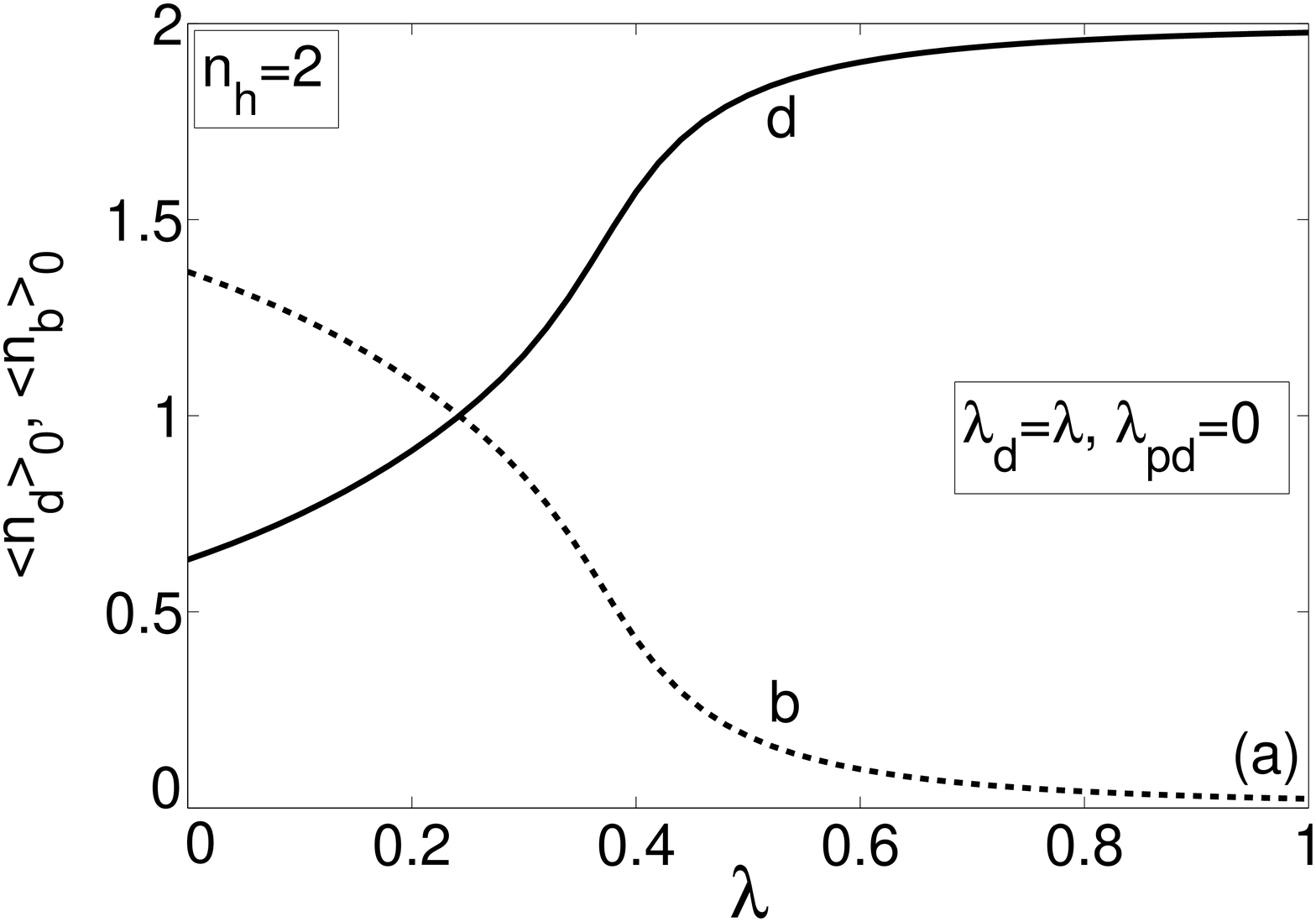}
%\hfill
\includegraphics[width=0.45\linewidth]{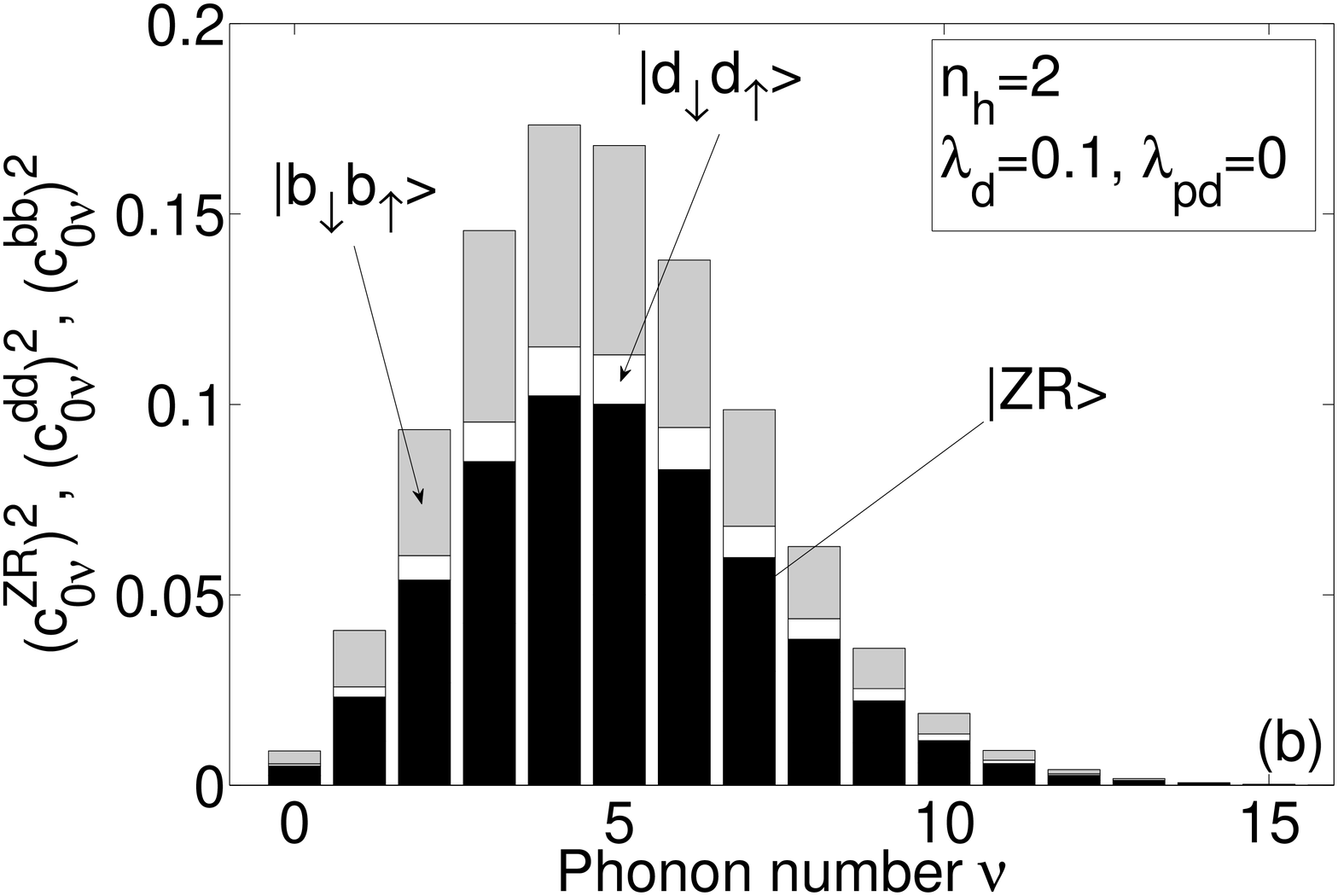}
%\hfill
\includegraphics[width=0.45\linewidth]{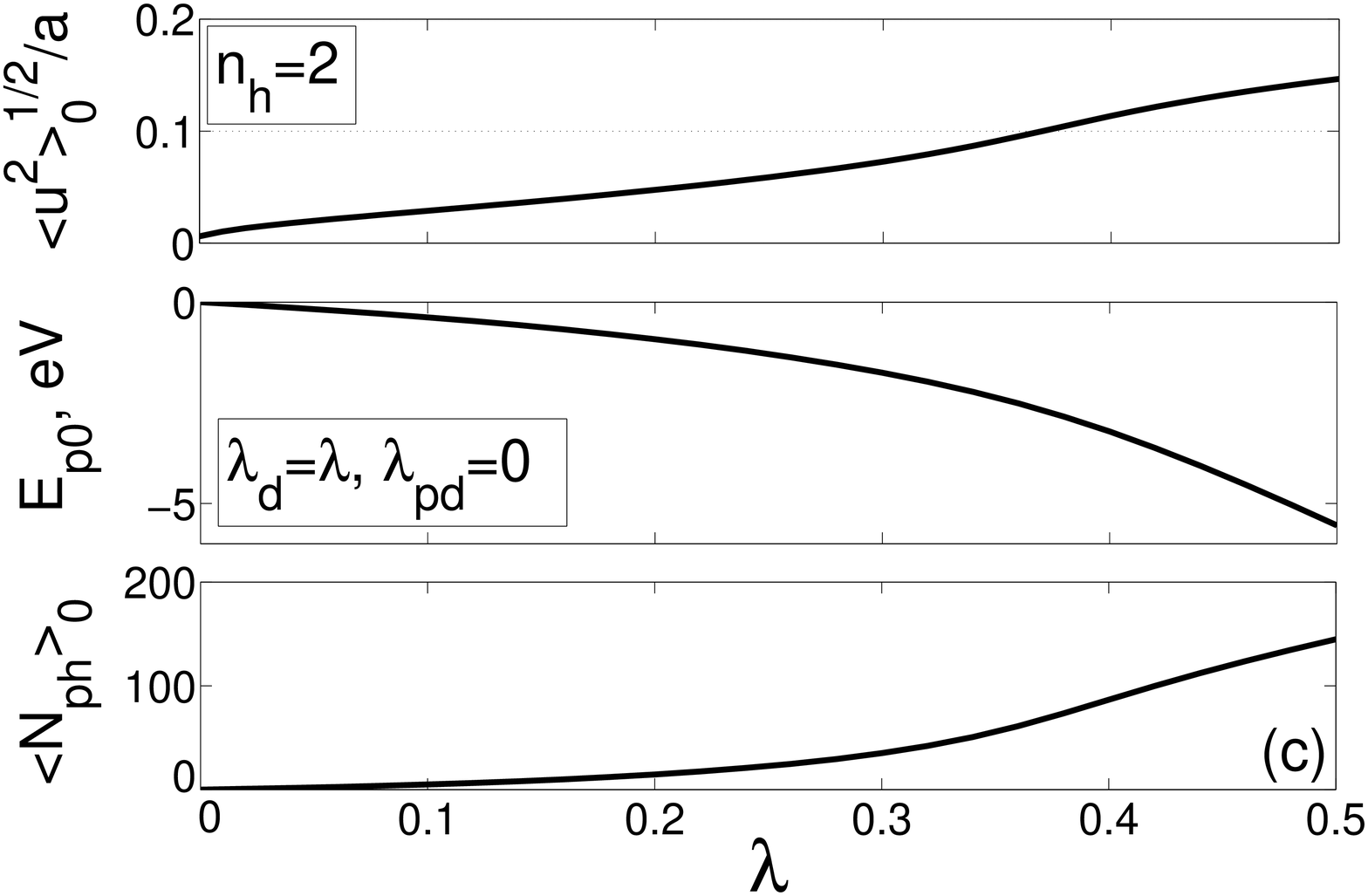}
%\hfill
\includegraphics[width=0.45\linewidth]{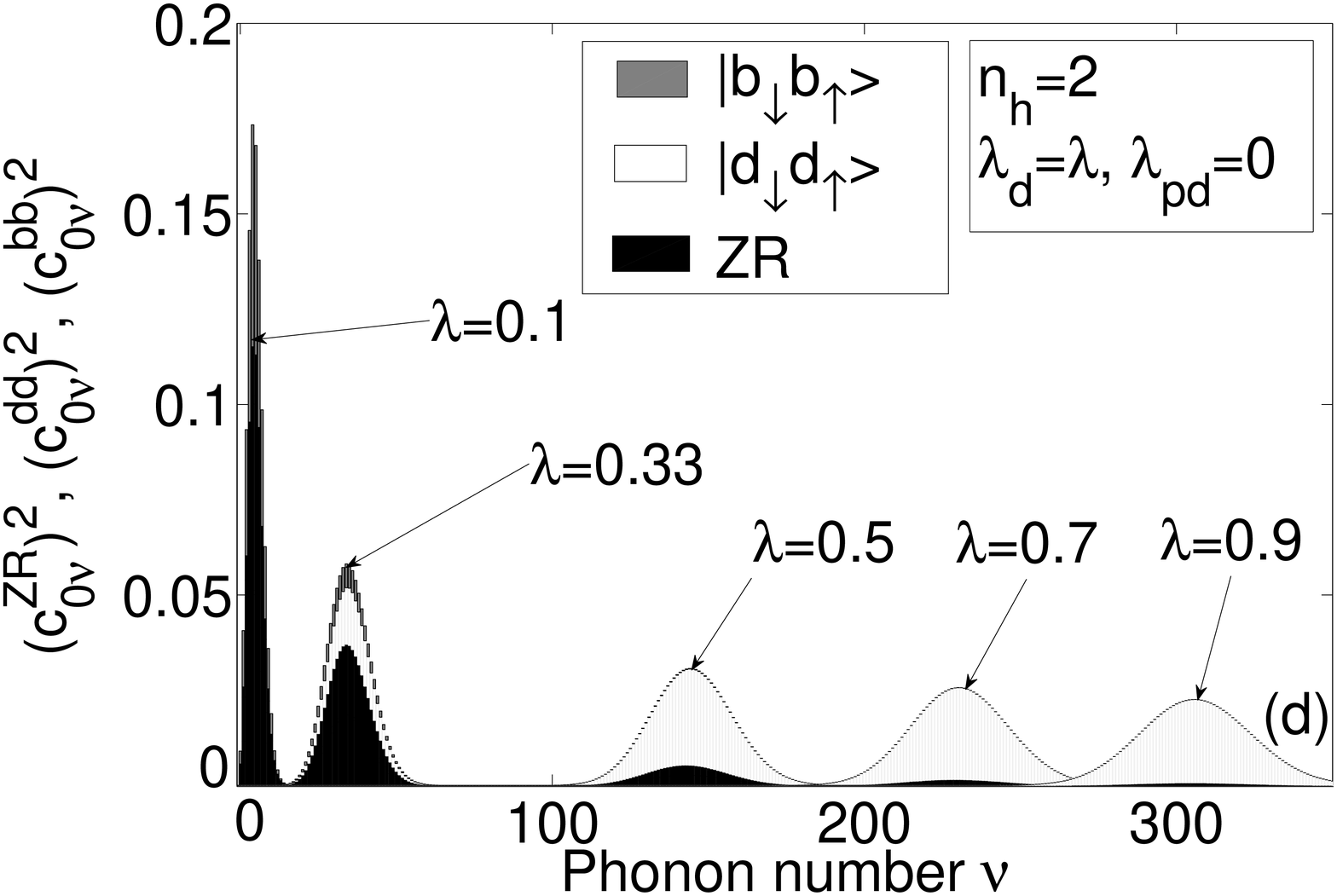}
%\hfill
\caption{\label{fig:fnh2_ld} The distribution of  the two holes among copper $d$ and oxygen $b$ orbitals as  a function of the diagonal EPI $\lambda$ and phonon number $\nu$ for (a) small and (b) large coupling, (c) the ground state ratio of the square root of the average square of oxygen displacement to the lattice parameter ${\frac{\sqrt {{{\left\langle {{u^2}} \right\rangle }_0}} }{a}}$, polaronic shift $E_{p0}$, and average number of phonons ${\left\langle {{N_{ph}}} \right\rangle _0}$ dependences on the diagonal EPI.}
\end{figure*}
\begin{figure}
\center
\includegraphics[width=1\linewidth]{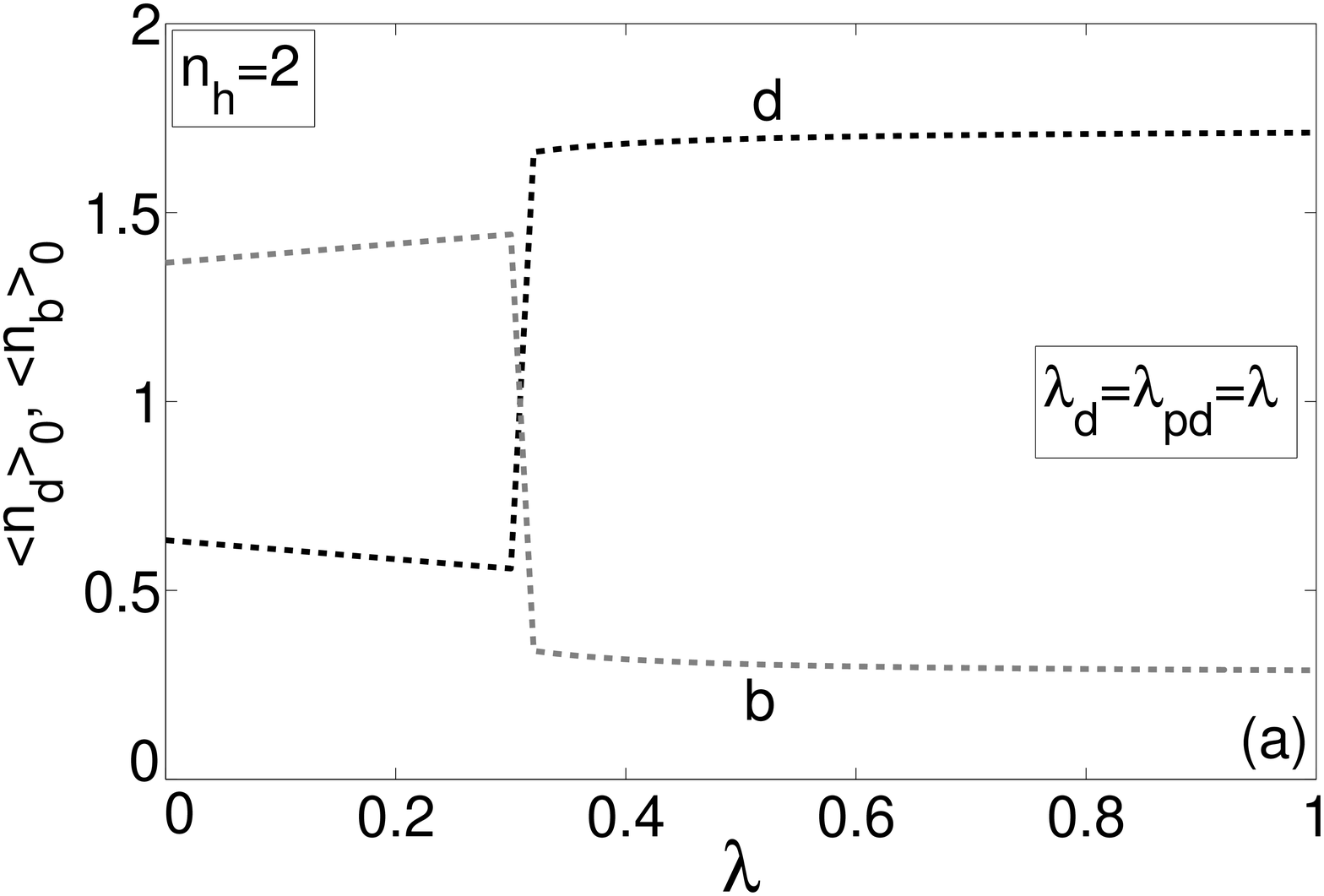}
\includegraphics[width=1\linewidth]{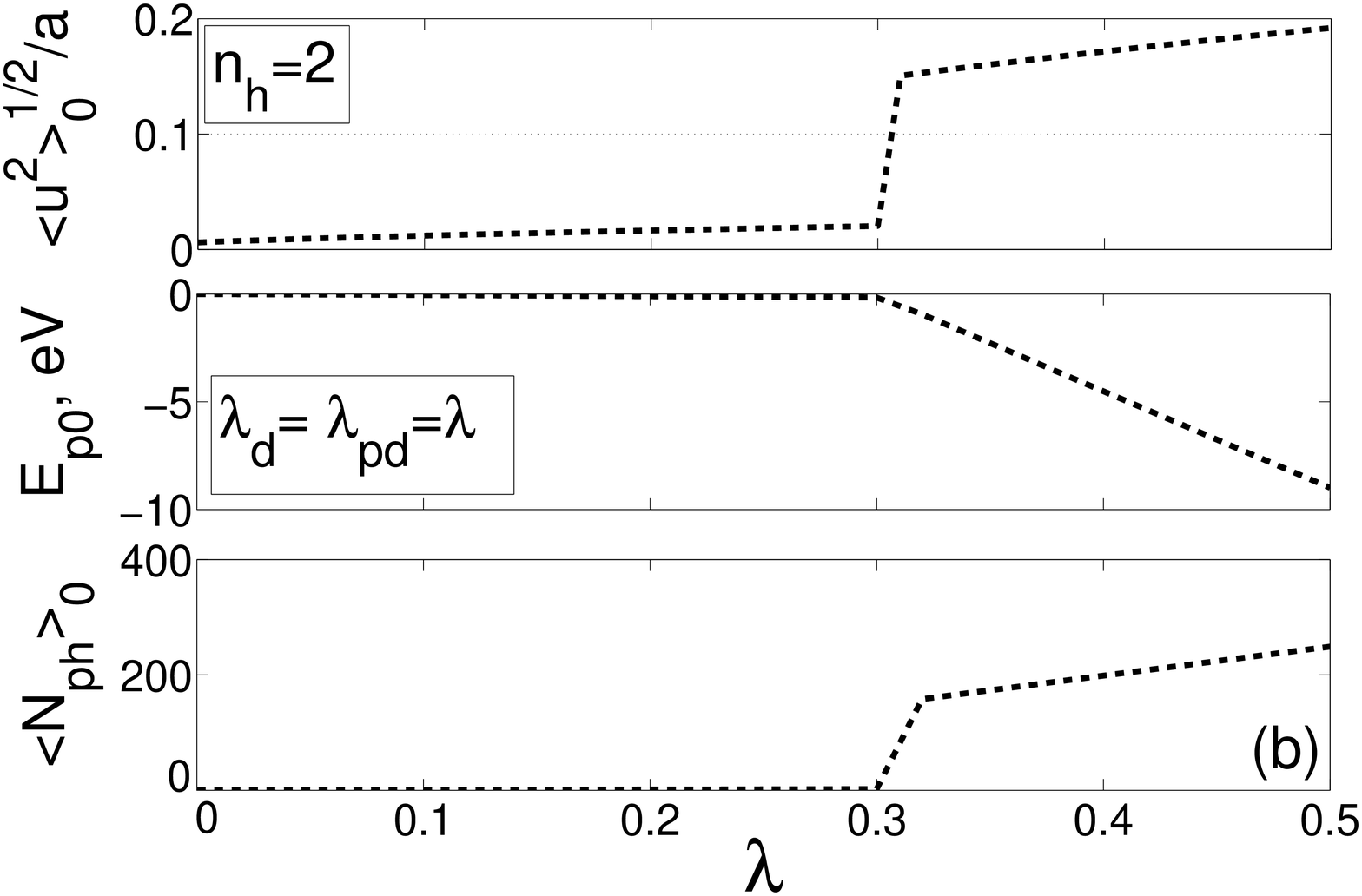}
\includegraphics[width=1\linewidth]{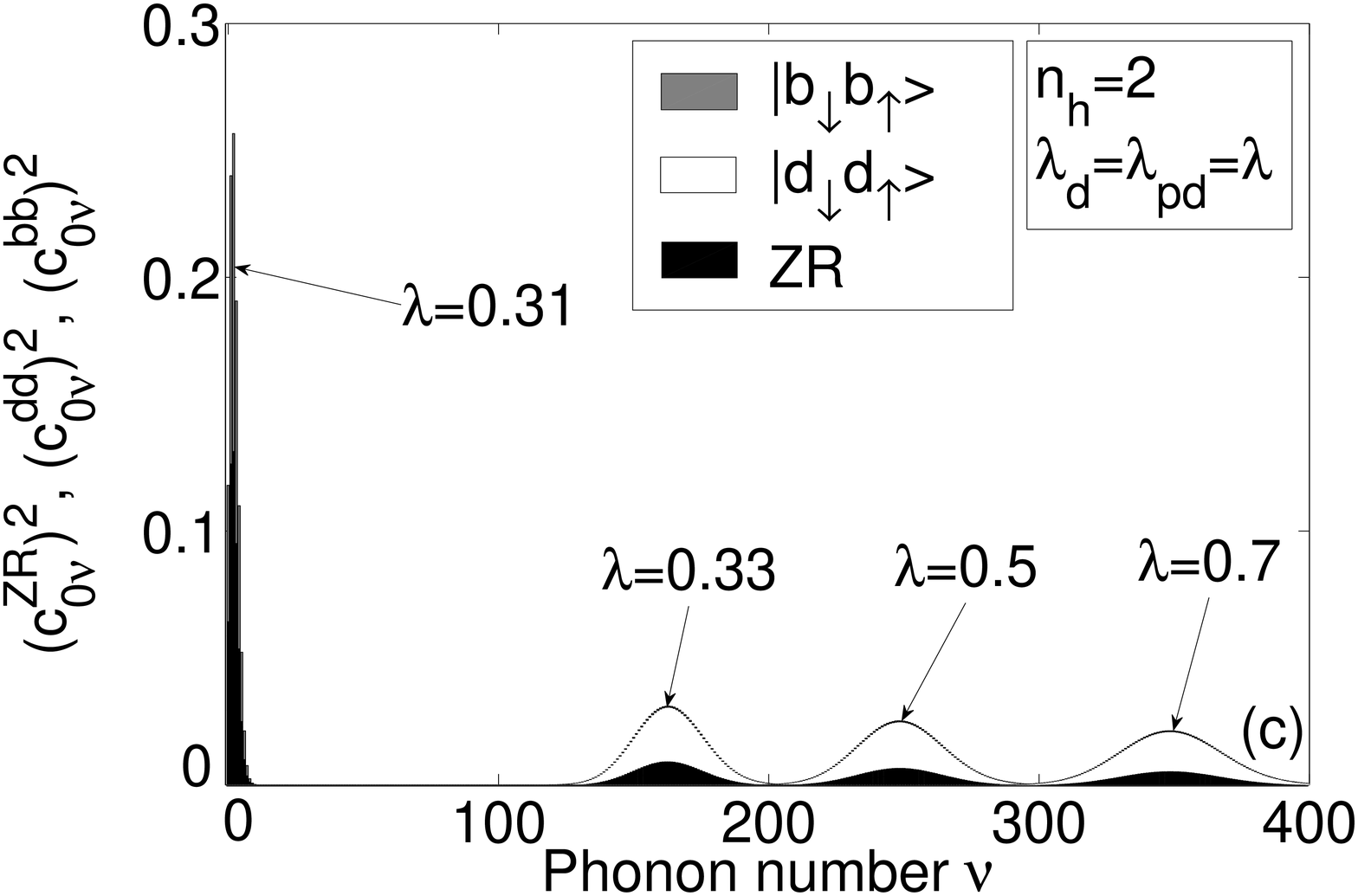}
\caption{\label{fig:fnh2_lpd} The effect of the equal diagonal and off-diagonal EPI on the distribution of the two holes as the function of (a) EPI and (c) phonon number, (b) the ground state ratio of the square root of the average square of oxygen displacement to the lattice parameter ${\frac{\sqrt {{{\left\langle {{u^2}} \right\rangle }_0}} }{a}}$, polaronic shift $E_{p0}$, and average number ${\left\langle {{N_{ph}}} \right\rangle _0}$ of phonons.}
\end{figure}
Two-hole states of the CuO$_4$ cluster may be singlets or triplets. For the low energy theory only singlets are essential. We write down the exact eigenstates in the form
\begin{eqnarray}
\label{es2}
\left| {2,j} \right\rangle & = & \sum\limits_{\nu  = 0}^N \left(  c_{j\nu }^{ZR}\left| {{\rm{ZR}}} \right\rangle \left| \nu  \right\rangle + \right. \nonumber \\
 &+ & \left. c_{j\nu }^{dd}\left| {{d_{\downarrow }}{d_{\uparrow }}} \right\rangle \left| \nu  \right\rangle  + c_{j\nu }^{bb}\left| {{b_ \downarrow }{b_ \uparrow }} \right\rangle \left| \nu  \right\rangle \right)
\end{eqnarray}
Here the first term corresponds to the Zhang-Rice (ZR) configuration $\left| {3{d^9}2{p^5}} \right\rangle $, the second one to the two holes on copper $\left| {3{d^8}2{p^6}} \right\rangle $, and the third one has two holes on oxygen $\left| {3{d^{10}}2{p^4}} \right\rangle $. The two-hole state appears in the theory due to electron removal process from the single-hole state. The additional hole occupy mainly the oxygen orbital. From Fig.~\ref{fig:fnh2_ld}a it is clear that in the absence of EPI, at ${\lambda _d} = \lambda_{pd} =0$ the number of copper holes has increased by $0.04$ in comparison to Fig.~\ref{fig:fnh1}a while the number of oxygen holes has grown almost by $1$. In the ground two-hole state the main contribution is given by the ZR contribution, the minimal weight has $\left| {{d_{\downarrow }}{d_{\uparrow }}} \right\rangle $ contribution, and the $\left| {{b_ \downarrow }{b_ \uparrow }} \right\rangle $ configuration has intermediate weight (Fig.~\ref{fig:fnh2_ld}b).

If electrons were free of the Coulomb interaction one would expect a doubling of many polaron characteristics vs single-hole state ones.  Due to the strong electron correlations in cuprates situation is different. Because the second hole occupies mainly the oxygen orbital the diagonal EPI results in a small changes vs single-hole eigenstates. For ${\lambda _d} < 0.1$ the average square of displacement ${\left\langle {u^2} \right\rangle}_0$, number of phonons $\left\langle {{N_{ph}}} \right\rangle $, and polaronic binding energy  increases less then $25\%$ in comparison to the single-hole eigenstates (Fig.~\ref{fig:fnh2_ld}c). Further increase of the diagonal EPI provides a smooth growth of the copper hole population, at ${\lambda_d} = 0.24$ it becomes equal to the oxygen hole population (Fig.~\ref{fig:fnh2_ld}a) and is close to $1$ at ${\lambda_d} = 0.9$. Thus we have found for two-hole eigenstates the effect of competition between the Coulomb repulsion and effective attraction mediated by EPI. Similar competition has been revealed earlier by the Quantum Monte Carlo method.~\cite{Johnston2013} The transformation of the phonon cloud with the maximum at $0$-phonon component to the multiphonon maximum for two-hole states~(\ref{es2}) with diagonal EPI occurs smoothly in the region ${\lambda_d} = 0.025 - 0.03$ (almost the same as for single-hole states~(\ref{es1})), nevertheless the evolution the large local polaron to the small local polaron continues up to ${\lambda _d} = 0.9$(Fig.~\ref{fig:fnh2_ld}d)

In the regime of equal diagonal and off-diagonal EPI up to $\lambda  < 0.314$ the population of oxygen holes negligibly increases and the copper holes population decreases with strengthening the EPI(Fig.~\ref{fig:fnh2_lpd}a).  At the critical point $\lambda  = 0.314$ there is a crossover between the different two-hole configurations, the maximal population is acquired by the $\left| {{d_{\downarrow }}{d_{\uparrow }}} \right\rangle $-configuration. The copper holes number increases sharply almost by $1$ (Fig.~\ref{fig:fnh2_lpd}a) at the critical point. Corresponding jumps have been revealed for the average square of displacement ${\left\langle {u^2} \right\rangle}_0$ and number of phonons $N_{max}$ (Fig.~\ref{fig:fnh2_lpd}b). The maximal number of phonons sharply changes from  $\nu = 15$ till $\nu = 190$ (Fig.~\ref{fig:fnh2_lpd}c).

Summarizing this chapter we plot schematically in Fig.~\ref{fig:hilbert_space}a the general structure of the relevant Hilbert space with three subspaces with the number of holes ${n_h} = 0,1,2$. With maximal phonon number $N_{\max}$ each electronic level is splitted into $\left( {N_{\max }} + 1 \right)$ sublevels. In the hole vacuum subspace there are $\left( {N_{\max }} + 1 \right)$ equidistant sublevels numerated by number of phonons and separated by a phonon energy $\hbar \omega _{br}$. In the single-hole subspace there are two blocks of spin doublets with the $2\left( {N_{\max } + 1} \right)$ sublevels each, corresponding to the bonding and antibonding hole orbitals. In the two-hole subspace there are three singlet blocks of $\left( {N_{\max } + 1} \right)$  sublevels and the triplet block of $3\left( {N_{\max } + 1} \right)$. In subspaces ${n_h} = 1$ and ${n_h} = 2$ the energy levels are not equidistant and the phonon number is not a quantum number.

Due to the long-range antiferromagnetic order in the undoped La$_2$CuO$_4$ with two sublattices $A$ and $B$ the local hole states from Eq.~(\ref{es1}) are subject of the effective exchange interaction $J \sim \frac{{{t^2}}}{U}$ that results in the splitting of states $\left| {1 \sigma ,i} \right\rangle $ and $\left| {1 \bar \sigma  ,i} \right\rangle $, here $\bar \sigma $ means $- \sigma $. From the chemical potential equation for undoped La$_2$CuO$_4$ at $T=0$ we found the only occupied term, that is the ${n_h} = 1$ ground state marked by cross in Fig.~\ref{fig:hilbert_space}a. According to the GTB approach (see Section~\ref{sec_pGTB} above) the electron removal from this initial state results in formation of the set of hole quasiparticles with different final states in the  subspace ${n_h} = 2$. These excitations are shown schematically by arrow in Fig.~\ref{fig:hilbert_space}a. Their dispersion is studied in the next Section~\ref{band_str}, all these excitations form the valence (v) band. Similarly, the electron addition to the occupied single-hole term results in formation of a set of hole vacuum multiphonon states~(\ref{es0}) and the conductivity (c) band.

%===========================================================================
\section{Polaronic band structure in the generalized tight binding method \label{band_str}}
%===========================================================================
\begin{figure*}
\center
\includegraphics[width=0.55\linewidth]{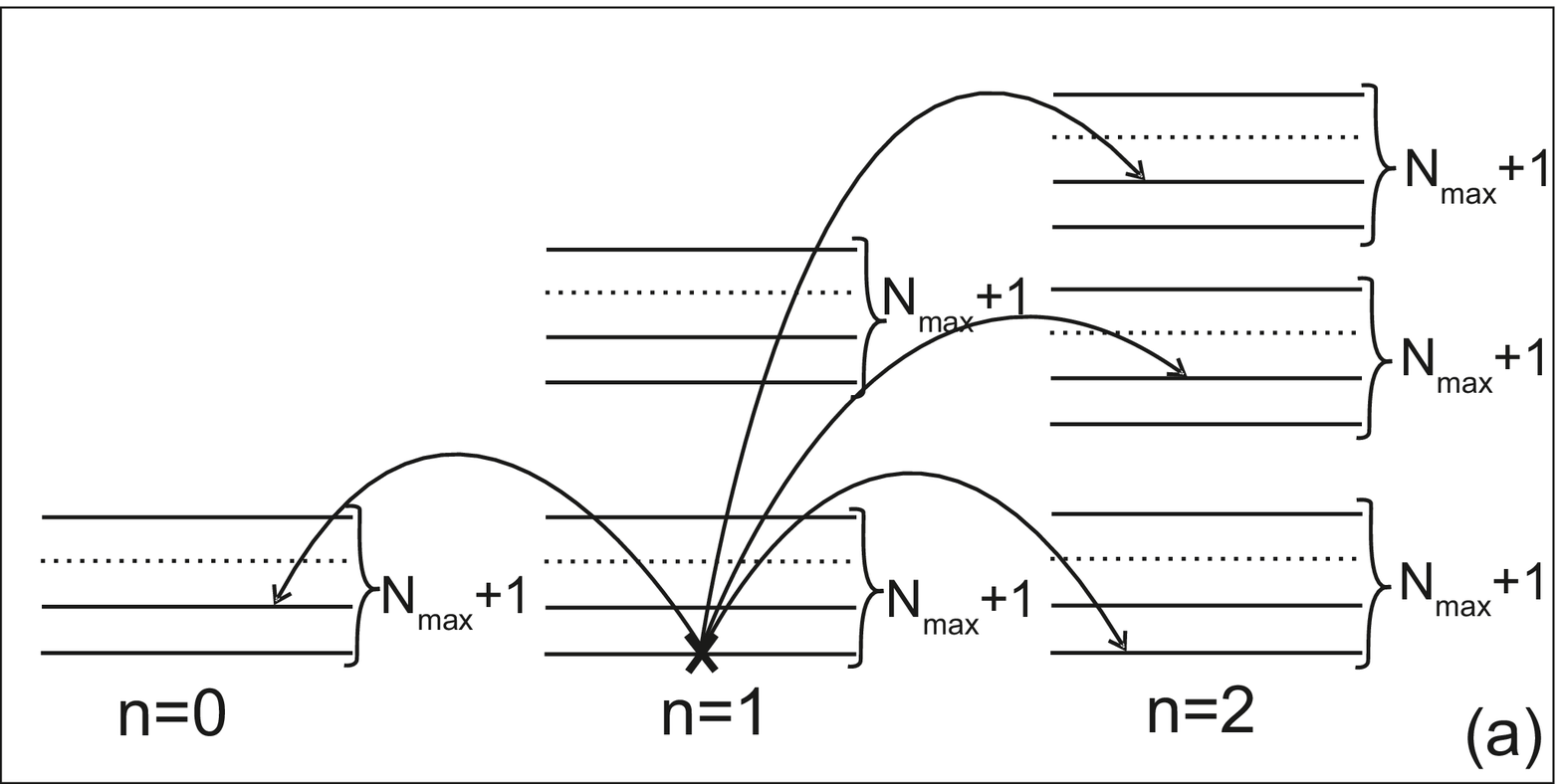}
\includegraphics[width=0.4\linewidth]{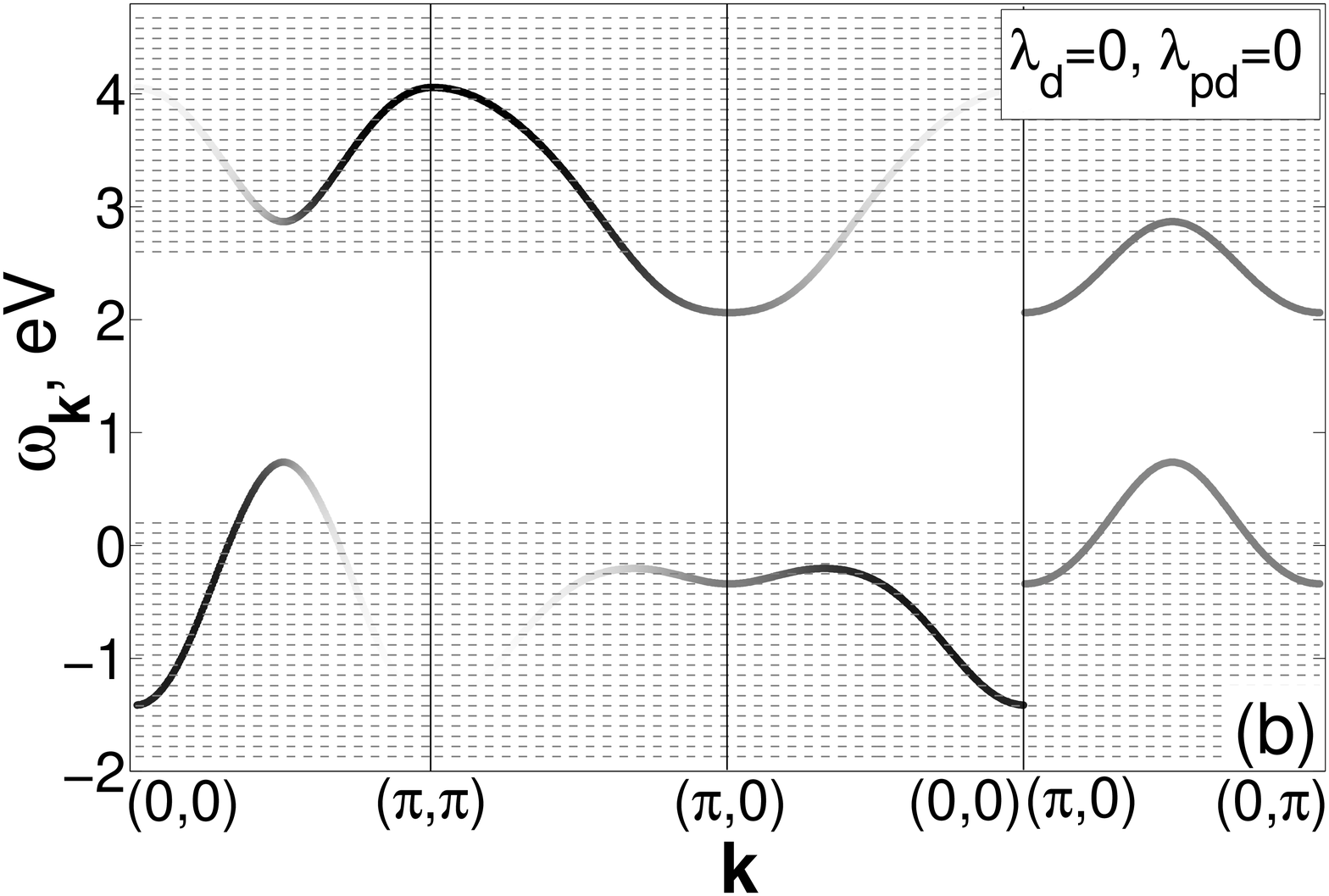}
\caption{\label{fig:hilbert_space} (a) The schematic picture of the multielectron and multiphonon local eigenstates for number of holes ${n_h} = 0,1,2$ per CuO$_4$ unit cell and maximal number of phonons  $N_{max}=2$. In the two-hole sector, the irrelevant triplet states with higher energy are not shown. Arrows between subspaces $\left( {0,1} \right)$ and $\left( {1,2} \right)$ schematically indicate polaronic quasiparticles. (b) Quasiparticles dispersion $\omega_{\bf{k}}$ without EPI.  Solid lines correspond to the conductivity and the valence bands of electrons in the antiferromagnetic phase. Dotted horizontal lines correspond to the dispersionless Franck-Condon resonances with zero spectral weight in the absence of EPI.}
\end{figure*}
\begin{figure*}
\includegraphics[width=0.45\linewidth]{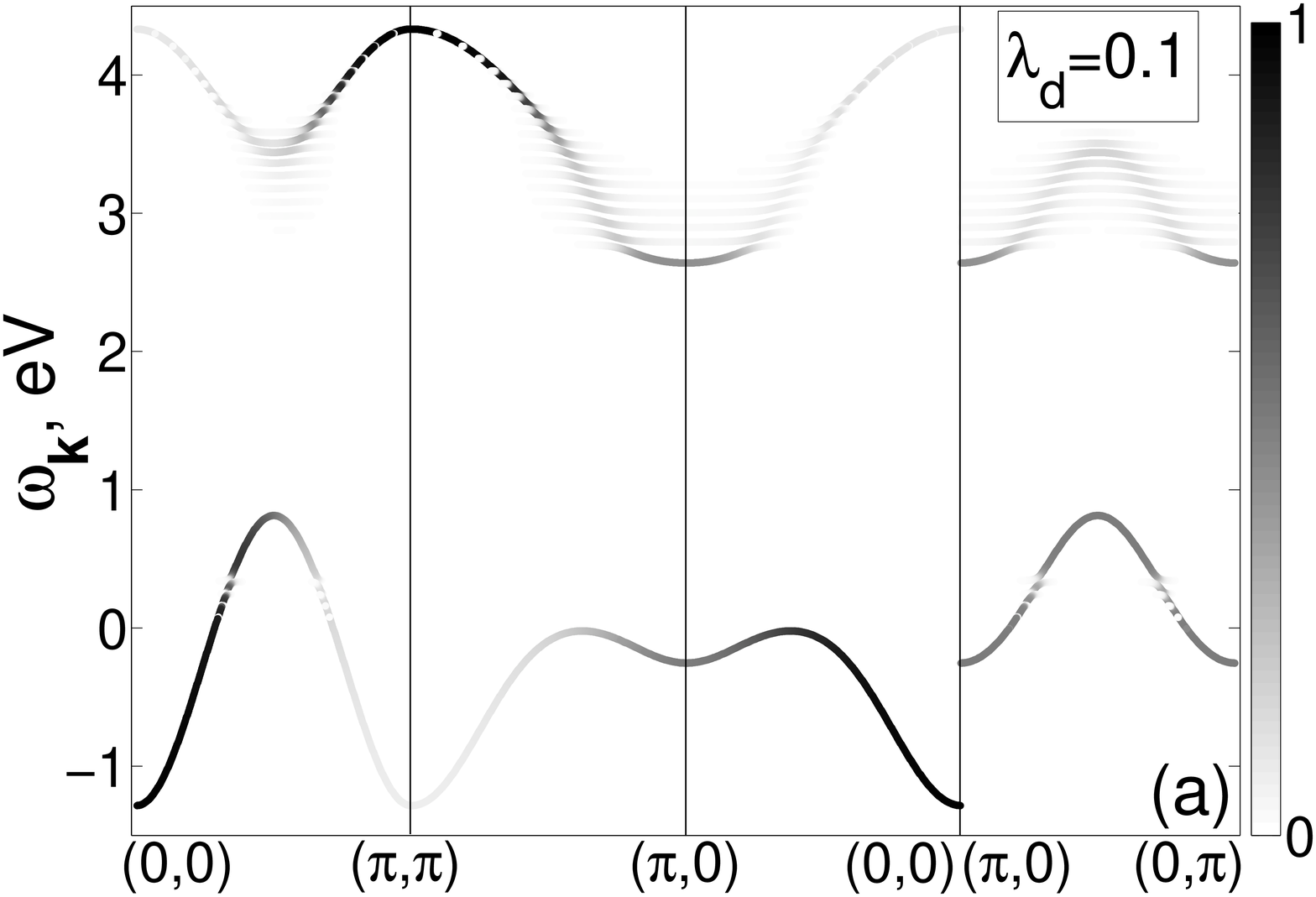}
\includegraphics[width=0.45\linewidth]{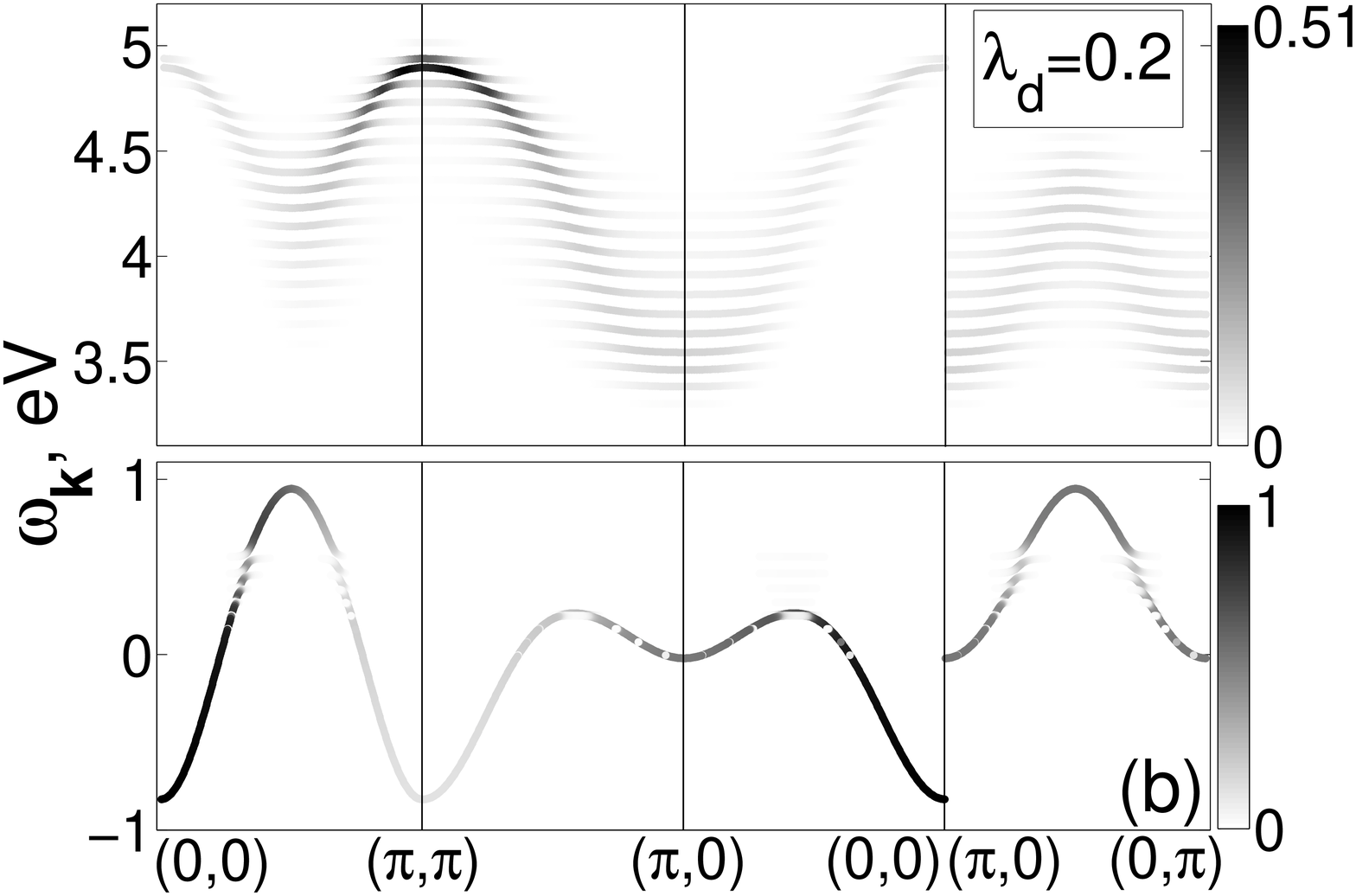}
\includegraphics[width=0.45\linewidth]{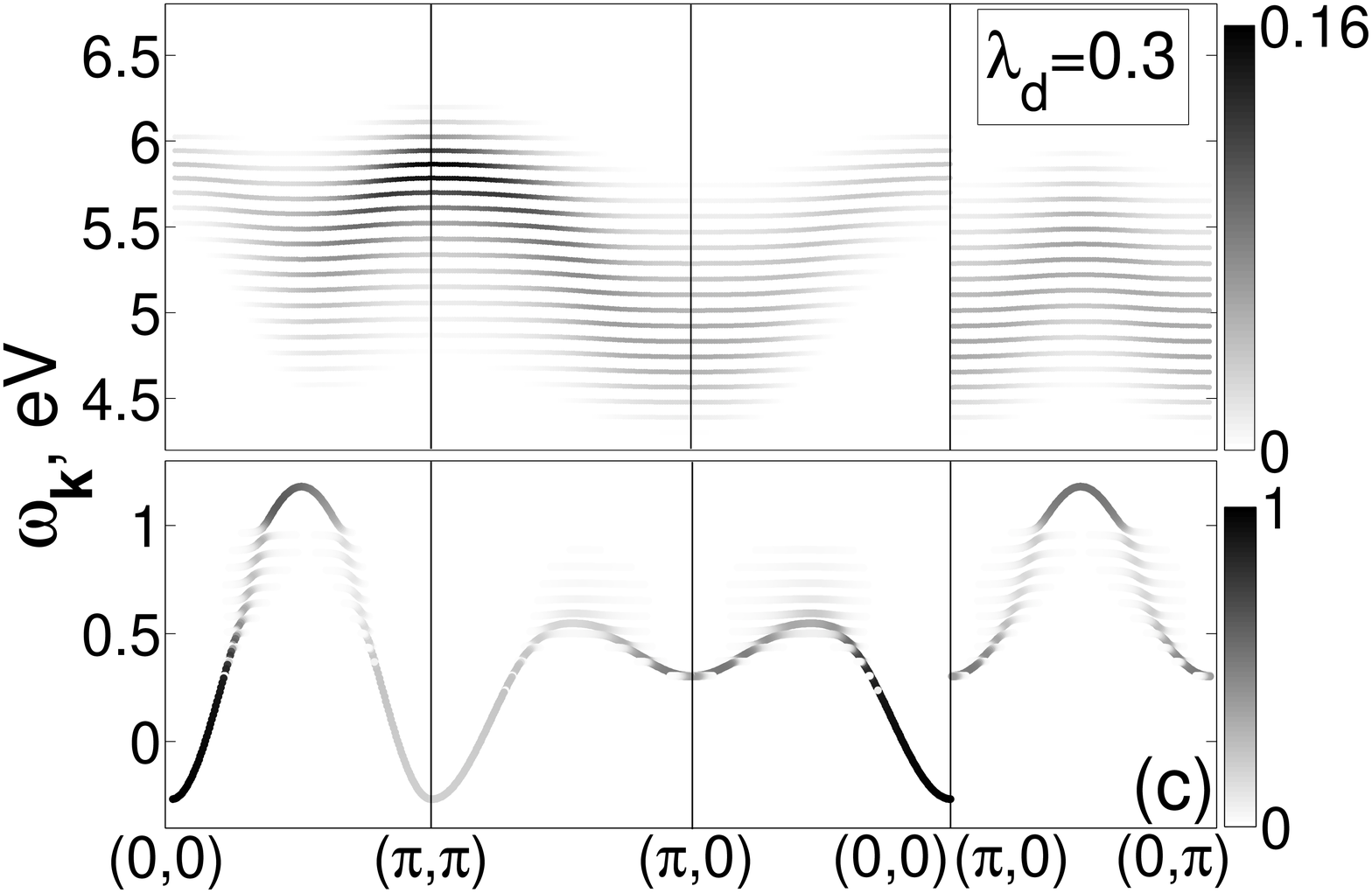}
\includegraphics[width=0.45\linewidth]{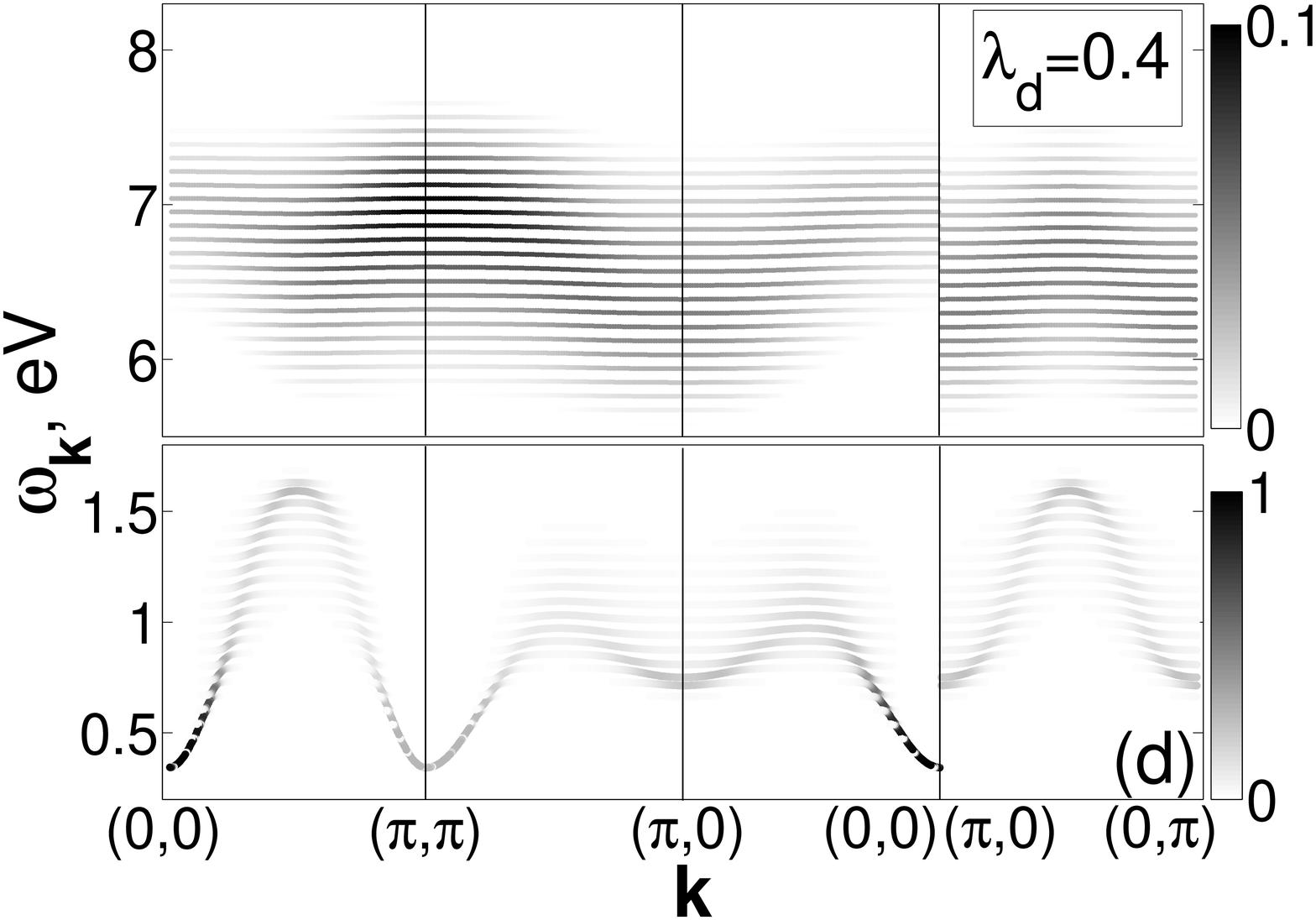}
\caption{\label{fig:bandstr_lamd} Effect of the diagonal EPI on the polaronic band structure for undoped antiferromagnetic La$_2$CuO$_4$ at $T = 10\,K$ for different values of EPI parameters: (a) ${\lambda _d} = 0.1$, (b) ${\lambda_d} = 0.2$, (c) ${\lambda_d} = 0.3$, (d) ${\lambda_d} = 0.4$. The line intensity is proportional to the quasiparticle spectral weight. We emphasize the decreasing of the intensity scale for the conductivity band with EPI growth.}
\end{figure*}
\begin{figure}
\includegraphics[width=0.9\linewidth]{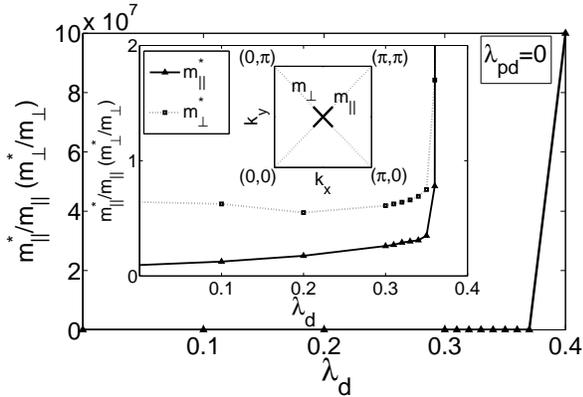}
\caption{\label{fig:d_effmass} The effective mass of the first removal state at the top of the valence band for different values of diagonal EPI, $\lambda_{pd}=0$.}
\end{figure}
\begin{figure*}
\includegraphics[width=0.45\linewidth]{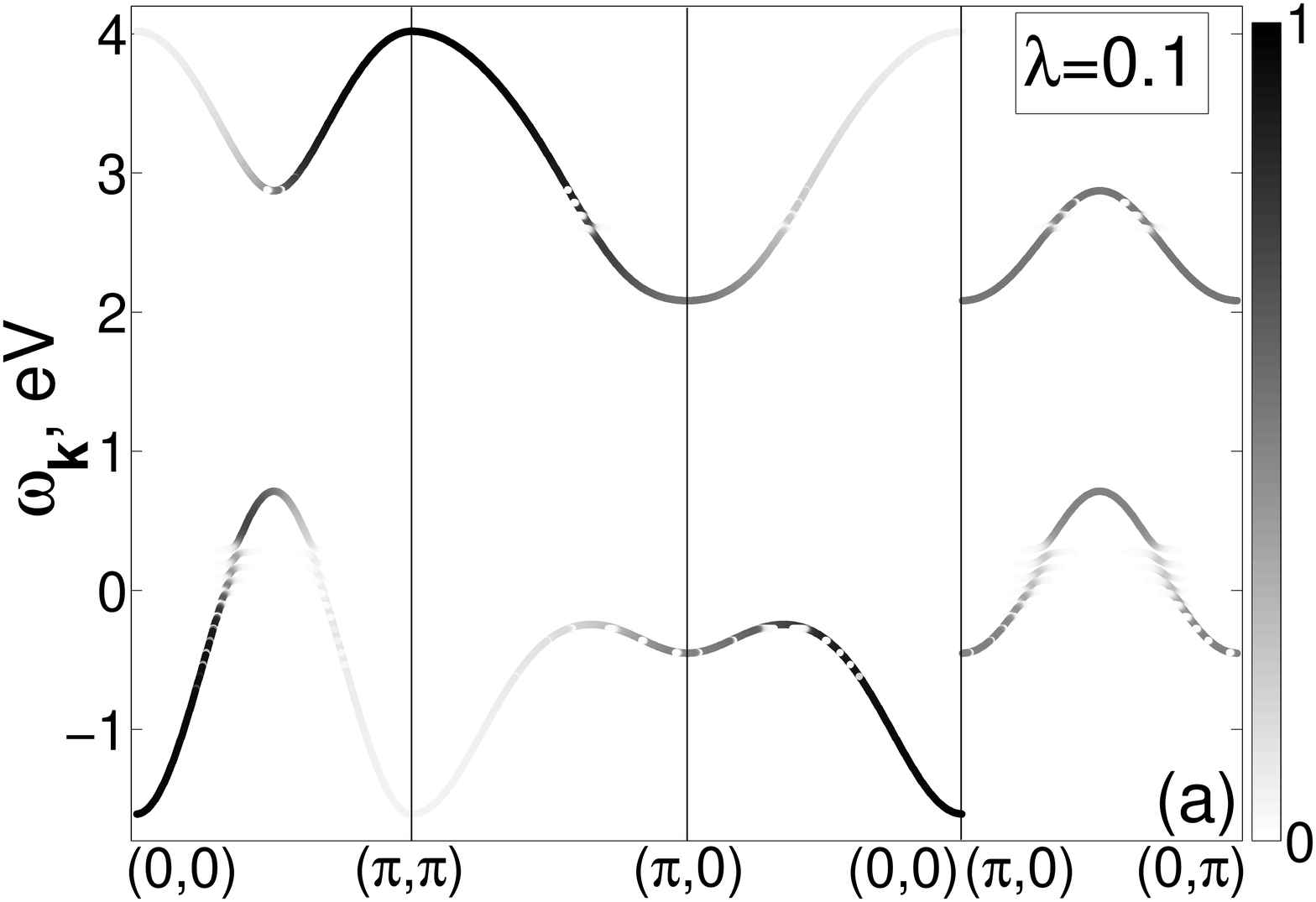}
\includegraphics[width=0.45\linewidth]{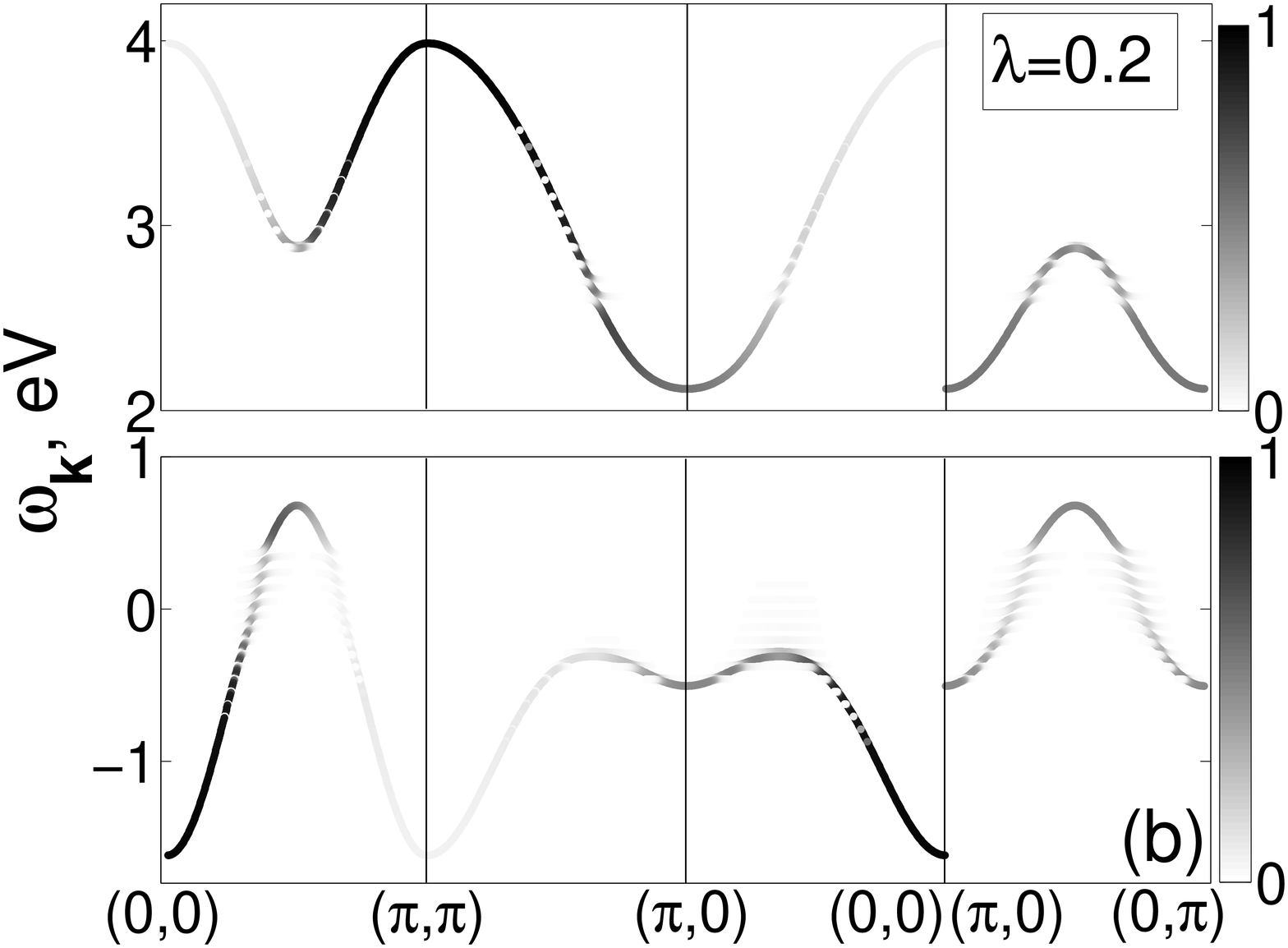}
\includegraphics[width=0.45\linewidth]{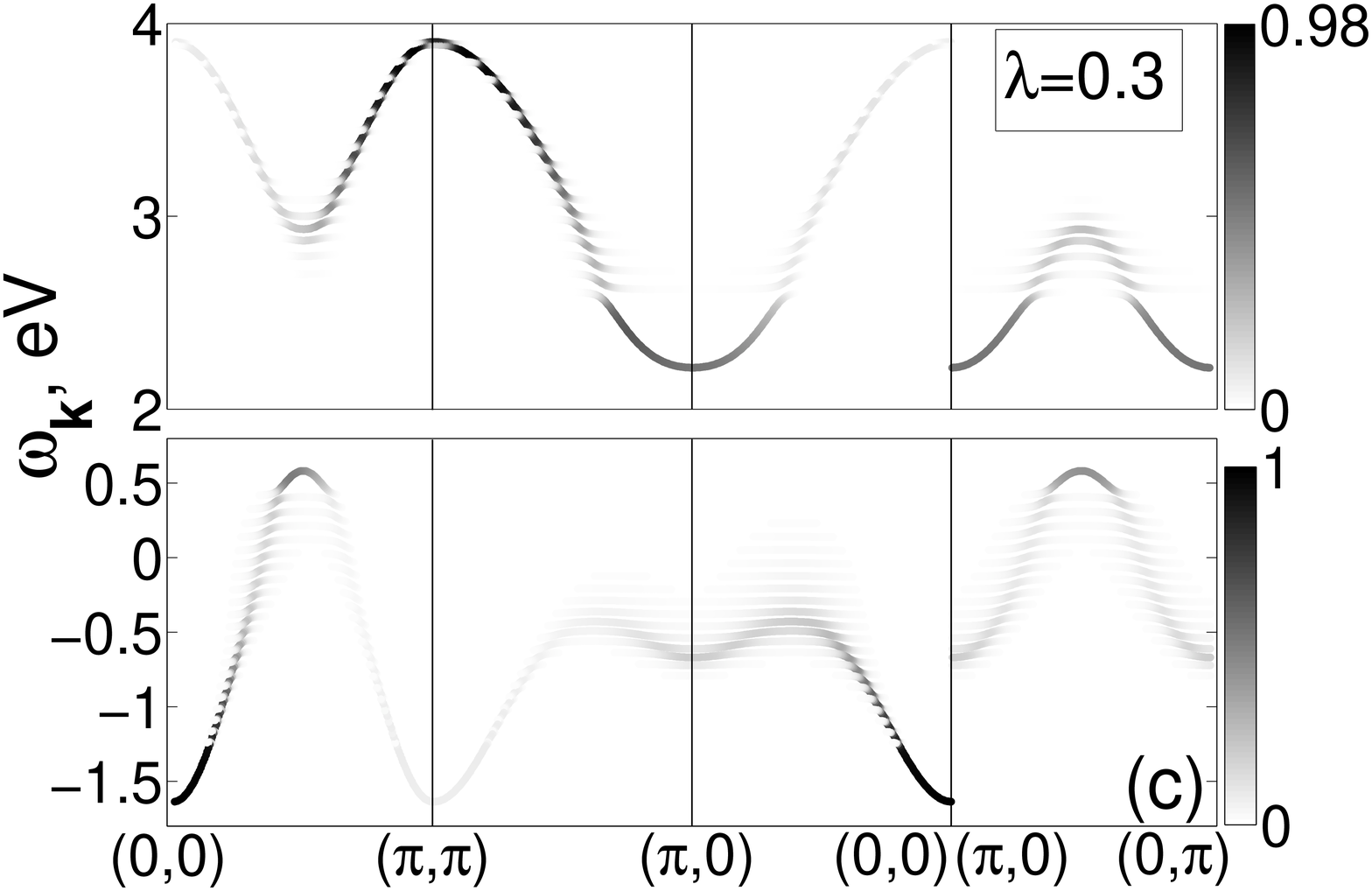}
\includegraphics[width=0.45\linewidth]{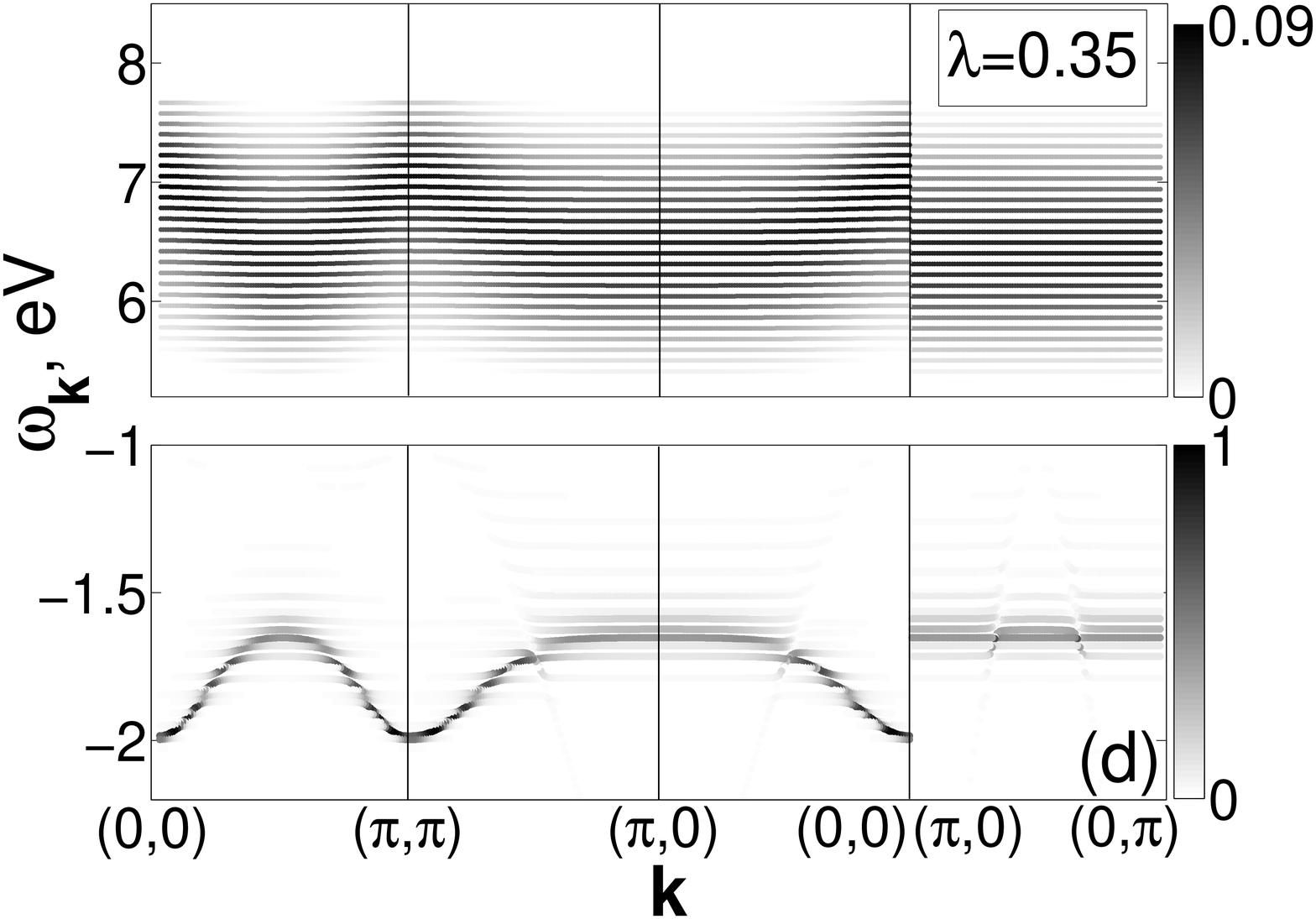}
\caption{\label{fig:bandstr_lampd} Effect of equal diagonal and off-diagonal EPI on the polaron band structure of the undoped antiferromagnetic La$_2$CuO$_4$ at $T = 10\,K$ for different values (a) ${\lambda} = 0.1$, (b) ${\lambda} = 0.2$, (c) ${\lambda} = 0.3$, (d) ${\lambda} = 0.35$.}
\end{figure*}
\begin{figure}
\includegraphics[width=0.9\linewidth]{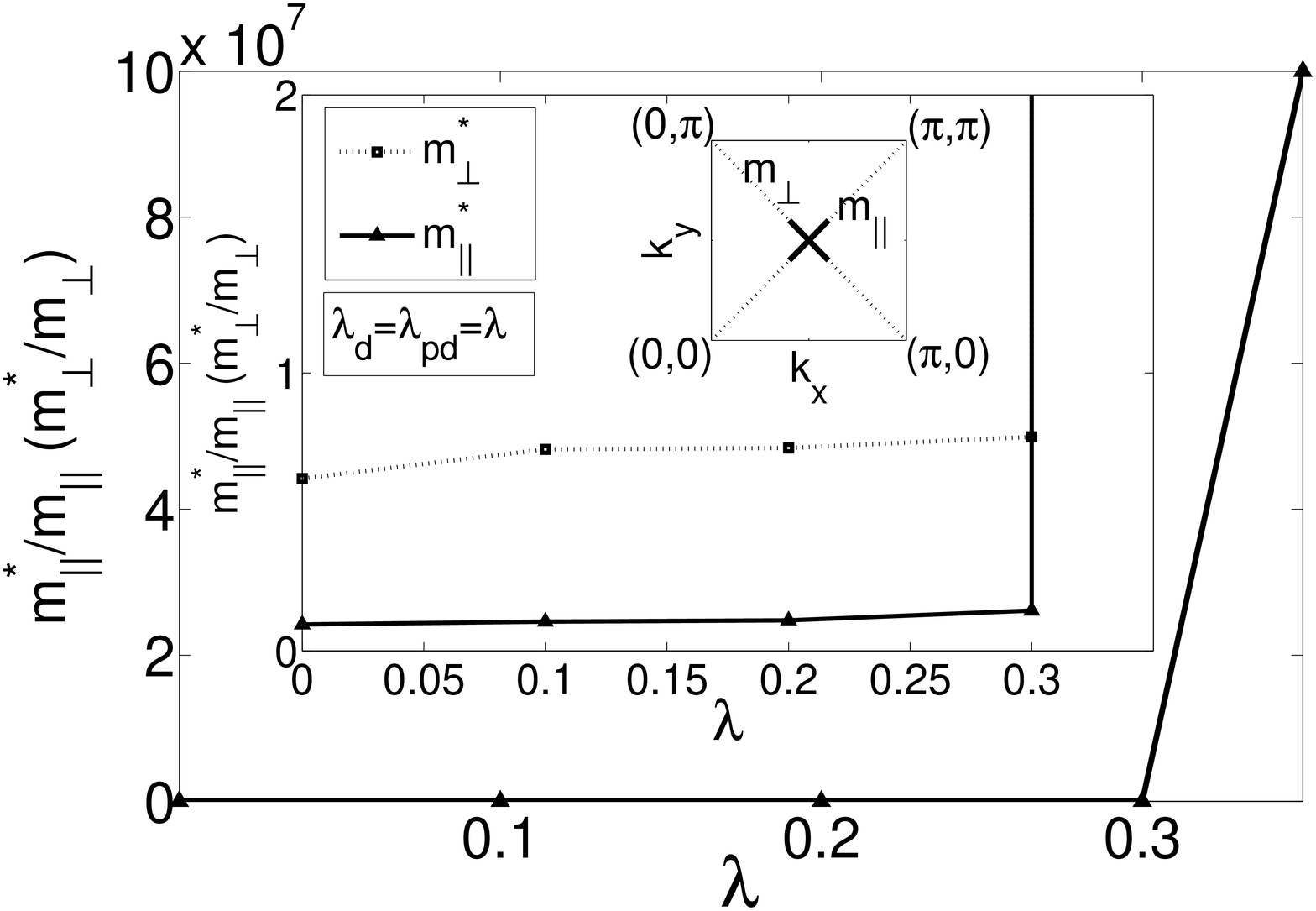}
\caption{\label{fig:dpd_effmass} The effective mass at the top of the valence band at $\left( {\frac{\pi }{2},\frac{\pi }{2}} \right)$ for equal diagonal and off-diagonal EPI.}
\end{figure}
The mathematical tool that allows to work with multilevel lattice system with a set of orthogonal and normalized local eigenstates $\left\{ {\left| p \right\rangle } \right\} = \left\{ {\left| ni \right\rangle } \right\} = \left\{ {|0,\nu \rangle ,\left| {1\sigma,i} \right\rangle ,\left| {2,j} \right\rangle } \right\}$ is given by the Hubbard operators ${X^{pq}} = \left| p \right\rangle \left\langle q \right|$.~\cite{Hubbard65} Each diagonal operator ${X^{pp}}$ determines the occupation of the eigenstate $\left| p \right\rangle $, while the nondiagonal operator $X_{\bf{f}}^{pq}$ describes the excitation at the site $\bf{f}$ from the initial state $\left| q \right\rangle $ to the final state $\left| p \right\rangle $. If the change of electric charge during the excitation is odd this excitation is the Fermi-type quasiparticle. According to the definition of our multielectron and multiphonon eigenstates, the Hubbard fermion in p-GTB is a polaron and we call it the Hubbard polaron. Due to completeness of the local set of eigenstates $\left\{ {\left| p \right\rangle } \right\}$  each single-hole annihilation operator at site $\bf{f}$ is given exactly by the linear combination of the Hubbard fermions.~\cite{Hubbard65}

We can write down the hole annihilation operators on corresponding orbitals $\beta = b,d$ as the linear combinations of the Hubbard fermions
\begin{eqnarray}
a_{\sigma \left( \beta \right) } = \sum\limits_{pq} {\gamma _{\sigma  \left( \beta \right)}({pq})} X_{\bf{f}}^{pq}
\label{fermi_op}
\end{eqnarray}
The matrix elements ${\gamma _{\sigma \left( \beta  \right)}}\left( {pq} \right) = \left\langle p \right|{a_{\sigma \left( \beta  \right)}}\left| q \right\rangle $
for orbital $\beta$ are calculated straightforwardly because we know all eigenstates $\left| p \right\rangle $  and $\left| q \right\rangle $. Quasiparticle transitions between states within one Hilbert space sector are described by Bose-type Hubbard operators $Z_{\bf{f}}^{ni,nj}$, to distingwish Bose operator from Fermi one we use for the former the notation Z.

The phonon number is not a quantum number in the single-hole~(\ref{es1}) and two-hole~(\ref{es2}) states therefore the phonon annihilation operator on the site $\bf{f}$ is given by
\begin{equation}
A_{\bf{f}}= \sum\limits_{n = 0}^2 {\sum\limits_{ij} \gamma _A \left( ni,nj \right) Z_{\bf{f}}^{ni,nj} }
 \label{Aphonon}
\end{equation}
where ${\gamma _A}\left( {{{ni}},{{nj}}} \right) = \left\langle {{{ni}}} \right|A_{\bf{f}}\left| {{{nj}}} \right\rangle $.

Since we know all coefficients in Eq.~(\ref{fermi_op}) and~(\ref{Aphonon}) it is easily to write down the total Hamiltonian $H_{c}+H_{cc}$ from Eq.~\ref{Hc_Hcc} in terms of Hubbard operators, then it consists of two parts. The first one, $H_{el-int}=H_c+H_{cc}^{el-el}$, is similar to the Hubbard model but constructed on operators describing Hubbard polaron excitations instead of usual Hubbard fermions ones. The second part, $H_{el-ph}=H_{cc}^{el-ph}$, results from EPI in initial Hamiltonian and contains Bose excitations.
\begin{widetext}
\begin{subequations} \label{H_X}
\begin{align}
\label{X_tot}
H_{tot} = H_c + H_{cc}, H_{cc}= H_{cc}^{el-el} + H_{cc}^{el-ph},%\\
\end{align}
\begin{align}
\label{X_Hc}
H_{c} = \sum\limits_{\bf{f}} { \sum\limits_{n = 0}^2 { \sum\limits_{i}{E_{ni}}X_{\bf{f}}^{ni,ni} }}, %\\
\end{align}%
\begin{align}%
\label{X_Hcc_el}
H_{cc}^{el-el} =-\sum\limits_{\bf{f,f'},\sigma,\beta} \sum\limits_{ii'jj'kk'll'} 2t_{p\beta}\mu_{\bf{ff'}} \left\{ \left( { \gamma_{\sigma \left(\beta\right)}^{*}\left( {0i,1\sigma j} \right) X_{\bf{f}}^{1\sigma j,0i} + \gamma_{\sigma \left(\beta\right)}^{*}\left( {1{\bar \sigma} k,2l} \right) X_{\bf{f}}^{2l,{1\bar \sigma}k}} \right)\times \right. \nonumber \\
\left. \left( { \gamma_{\sigma \left(b\right)}\left( {0i',1\sigma j'} \right) X_{\bf{f'}}^{0i',1\sigma j'} + \gamma_{\sigma \left(b \right)}\left( {1{\bar \sigma} k',2l'} \right) X_{\bf{f'}}^{{1\bar \sigma}k',2l'}} \right) + h.c. \right\}, %\\
\end{align}
\begin{align}
\label{X_Hcc_epi}
H_{cc}^{el-ph} = \sum\limits_{\bf{f,f'},\sigma} \sum\limits_{n,i,j}\sum\limits_{k,l,r}2\lambda_{d}\mu_{\bf{ff'}} \left( { \gamma_{\sigma \left(A\right)}^{*}\left( {ni,nj} \right) X_{\bf{f}}^{nj,ni} + \gamma_{\sigma \left(A \right)}\left( {ni,nj} \right) X_{\bf{f}}^{ni,nj}} \right)\times \nonumber \\
\left( { {\left| \gamma_{\sigma \left(d\right)} \left( {0l,1\sigma k} \right)\right|}^2 X_{\bf{f'}}^{1\sigma k,1\sigma k} + {\left| \gamma_{\sigma \left(d\right)}\left( {1{\bar \sigma} k,2r} \right)\right|}^2 X_{\bf{f'}}^{2r,2r}} \right) + \nonumber \\
+ \sum\limits_{\bf{f,f',h},\sigma} \sum\limits_{n,i,j}\sum\limits_{k,k',l,l'} 2 \lambda_{pd} \rho_{\bf{ff'h}}^A \left( { \gamma_{\sigma \left(A\right)}^{*}\left( {ni,nj} \right) X_{\bf{f}}^{nj,ni} + \gamma_{\sigma \left(A \right)}\left( {ni,nj} \right) X_{\bf{f}}^{ni,nj}} \right)\times \nonumber \\
\left\{ \left( { \gamma_{\sigma \left(d\right)}^{*}\left( {0k,1\sigma k'} \right) X_{\bf{f'}}^{1\sigma k',0k} + \gamma_{\sigma \left(d \right)}^{*}\left( {1{\bar \sigma} l,2l'} \right) X_{\bf{f}}^{2l',1{\bar \sigma}l}} \right)\times \right. \nonumber \\
\left. \left( { \gamma_{\sigma \left(b\right)}\left( {0l,1\sigma l'} \right) X_{\bf{h'}}^{0l,1\sigma l'} + \gamma_{\sigma \left(b \right)}\left( {1{\bar \sigma} k,2k'} \right) X_{\bf{h'}}^{{1\bar \sigma}k,2k'}} \right) + h.c. \right\}.
\end{align}
\end{subequations}
\end{widetext}
Here the intracluster term $H_c$ contains only energies ${E_{ni}}$ of the local eigenstates of the cluster with diagonal operators $X_{\bf{f}}^{{{ni}},{{ni}}}$, where $n=0,1,2$, and index $i$ enumerates the polaronic eigenstate in the Hilbert space sector with hole number $n$. The intercluster terms result in the polaronic hopping and dispersion. Due to the intercluster contribution from the diagonal and off-diagonal EPI the polaron scattering $H_{cc}^{el-ph}$ on the bosonic excitations occurs. In the intercluster matrix elements the single electron hopping $t_{pd}$, $t_{pp}$ and EPI $\lambda_d$, $\lambda_{pd}$ parameters are strongly suppressed by the matrix elements from Eq.~(\ref{fermi_op}) and~(\ref{Aphonon})  and structural factors ${\mu _{\bf{fg}}}$, ${\nu _{\bf{fg}}}$, ${\rho _{\bf{fgh}}}$. These factors are strongly decreasing with distance, and usually only contributions from the first three neighbors are enough, that is why the $t - t' - t''$-tight binding fitting is rather successful in cuprates.~\cite{Gav2000}

To obtain dispersion of quasiparticle excitations we use the method of equation of motion for the two-sublattice matrix Green function of polarons $D_{\bf{ff'}}^{GG'}\left( {uv;v'u'} \right) = \left\langle {\left\langle {{X_{{\bf{f}}G}^{uv}}}  \mathrel{\left | {\vphantom {{X_{{\bf{f}}G}^{uv}} {X_{{\bf{f'}}G'}^{v'u'}}}}
 \right. \kern-\nulldelimiterspace} {{X_{{\bf{f'}}G'}^{v'u'}}} \right\rangle } \right\rangle $. Due to the large number of fermionic quasiparticles it is convenient to introduce the matrix Green function $\hat D_{\bf{ff'}} = \left\{ {{D_{\bf{ff'}}^{mn}}} \right\}$, where row $m$ and column $n$ indexes numerate the different quasiparticles ${\left( {p,q} \right)_{n\left(m\right)}} \leftrightarrow n\left(m\right)$. Indeed, the number of quasiparticles ${\left( {p,q} \right)}$ is finite, so each may be enumerated just by the number $n\left(m\right)$ which has a meaning of the quasiparticle band index. Total Green function in the matrix form looks like
\begin{eqnarray}
{\hat D_{\bf{ff'}}} = \left( {\begin{array}{*{20}{c}}
{\hat D_{{\bf{ff'}}}^{AA}}&{\hat D_{{\bf{ff'}}}^{AB}}\\
{\hat D_{{\bf{ff'}}}^{BA}}&{\hat D_{{\bf{ff'}}}^{BB}}
\end{array}} \right)
\label{matrix_GF}
\end{eqnarray}
For the Hamiltonian with two-particle electron-electron interactions the generalized Dyson equation for the matrix Green function~\cite{OvchinnikovValkov} in the Fourier transformation reads
\begin{equation}
\hat D_{\bf{k}}\left(\omega \right) = {\left[ {\hat G_0^{ - 1}\left(\omega \right) - \hat P_{\bf{k}}\left(\omega \right){{\hat t}_{\bf{k}}} + \hat \Sigma_{\bf{k}} \left(\omega \right)} \right]^{ - 1}}\hat P_{\bf{k}}\left(\omega \right)
\label{gen_Dyson}
\end{equation}
Here $\hat G_0^{ - 1}(\omega )$ is a local propagator determined by the multielectron eigenstates $|p\rangle $ and $|q\rangle $, $\hat t_{\bf{k}}^{mn} = \gamma (m)\gamma (n){t_{\bf{k}}}$ is the intersite hopping matrix, and ${t_{\bf{k}}}$ is a bare dispersion. Besides the self-energy $\hat \Sigma ({\bf{k}},\omega )$ the unusual strength operator $\hat P({\bf{k}},\omega )$ appears in Eq.~(\ref{gen_Dyson}). It results in the redistribution of the QP spectral weight (or oscillator strength) and in renormalization of quasiparticle dispersion, that are the intrinsic features of SCES. Recently the ARPES line shape had been discussed with Eq.~(\ref{gen_Dyson}) and the odd in $\left( {{\bf{k}} - {{\bf{k}}_{Fermi}}} \right)$ contribution to the momentum distribution curve had been found due to the imaginary part of the strength operator.~\cite{OvchShnKord2014}
%In conventional Dyson equation for a single electron Green function a self-energy is determined by both contributions from the self-energy $\hat \Sigma_{\bf{k}}\left(\omega \right)$ and the strength operator $\hat P_{\bf{k}}\left(\omega \right)$ from Eq.~(\ref{gen_Dyson}).
% I think it should be rewritten.

The simplest nontrivial solution (Hubbard I approximation) is usually used in the cluster perturbation theory~\cite{SenechalPerPlouf2002,TremblayKyuSen2006}. It can be obtained from Eq.~(\ref{gen_Dyson}) when $\hat \Sigma _{\bf{k}}\left(\omega \right)  = 0$ and $\hat P_{\bf{k}}\left(\omega \right) = {\delta _{pq}}F\left( {pq} \right)$. The so-called filling factor $F\left( {pq} \right)$ is given by the sum of the initial and final state occupation numbers ${F_{\bf{f}}}\left( {pq} \right) = \left\langle {X_{\bf{f}}^{pp}} \right\rangle  + \left\langle {X_{\bf{f}}^{qq}} \right\rangle $ and is strongly dependent on sublattice magnetization, doping, and temperature. The appearance of the filling factor is very important difference of the Hubbard fermions vs bare electrons. In particular, this factor makes irrelevant many excitations from empty $\left| p \right\rangle $ to empty $\left| q \right\rangle $  states with determined energy ${E_p} - {E_q}$.

In the Hubbard I approximation we have obtained the following Dyson equation for the polaronic matrix Green function
\begin{equation}
\hat D_{\bf{k}}^{ - 1} = \hat D_0^{ - 1} + {\hat t_{\bf{k}}} + \hat M_{\bf{k}}^{EPI}.
\label{DysonEq}
\end{equation}
Here $D_0^{pq} = {{F\left( {pq} \right)} \mathord{\left/ {\vphantom {{F\left( {pq} \right)} {\left( {\omega - \Omega \left( {pq} \right)} \right)}}} \right.  \kern-\nulldelimiterspace} {\left( {\omega - \Omega \left( {pq} \right)} \right)}}$, $\Omega \left( {pq} \right)$ is the local quasiparticle energy of the multielectron and multiphonon eigenstates and ${\hat t_{\bf{k}}}$ is the matrix of combined $p-d$ and $p-p$ hopping that provides the band dispersion $\omega_{\bf{k}}$ in the absence of EPI.~\cite{Gav2000}

Matrix $\hat M_{\bf{k}}^{EPI}$ contains terms of intercluster EPI $\hat M_{\left( {pd} \right){\bf{k}}}^{\left( 1 \right)}$ and $\hat M_{\left( {pd} \right){\bf{k}}}^{\left( 2 \right)}$ which are presented in the Appendix~(\ref{link:7}). For the typical EPI parameters we have estimated the ratio $\frac{{{M_{pd}}}}{t}\sim 0.01$, and the terms $\hat M_{\left( {pd} \right){\bf{k}}}^{\left( 1 \right)}$ and $\hat M_{\left( {pd} \right){\bf{k}}}^{\left( 2 \right)}$ provide small contribution to the dispersion. Thus, the local intracluster effects of EPI discussed in the Section~\ref{exact_diag} are the most important for the polaronic bands formation.
In the absence of EPI the only polaronic quasiparticles with non-zero spectral weight are possible between the multielectron $n$, $n+1$ terms with equal number of phonons. At zero temperature this condition results only in $0 - 0$ Franck-Condon resonances between terms with the phonon number $\nu  = 0$. These Hubbard fermions dispesion is shown in Fig.~\ref{fig:hilbert_space}b by a solid line with different spectral weight in various parts of the Brilloin zone due to long range antiferromagnetic order.  All other polaronic excitations corresponds to the dispersionless Franck-Condon resonances with zero spectral weight in the absence of EPI (Fig.~\ref{fig:hilbert_space}b).

In general, the EPI results in the hybridization of the Hubbard subbands and the Franck-Condon resonances. The low energy part of the polaronic bands is shown in Fig.~\ref{fig:bandstr_lamd} and Fig.~\ref{fig:bandstr_lampd}. Effect of the diagonal and both diagonal and off-diagonal EPI on the polaronic band structure is different. Weak diagonal EPI introduced in~(\ref{Hc_Hcc}) mainly modifies the conductivity band while equal diagonal and off-diagonal coupling more strongly effect on the valence band. Thus at $\lambda  = 0.1$ we found the hybridization splitting and finite spectral weight for several Franck-Condon resonances in the conductivity band (Fig.~\ref{fig:bandstr_lamd}a). The number of splitted subbands and the value of the minigaps between these subbands becomes larger for $\lambda  = 0.2$ (Fig.~\ref{fig:bandstr_lamd}b). For the valence band there are also splitted subbands separated by smaller minigaps. The top of the valence band, particularly the first removal state at the $\left( {{\pi  \mathord{\left/
 {\vphantom {\pi  2}} \right.
 \kern-\nulldelimiterspace} 2},{\pi  \mathord{\left/
 {\vphantom {\pi  2}} \right.
 \kern-\nulldelimiterspace} 2}} \right)$, is unaffected by small diagonal EPI. For larger EPI the spectral weight of the conductivity band is strongly suppressed, it is transferred to the higher energy bands shown in Fig.~\ref{fig:hilbert_space}a. For ${\lambda _d} = 0.4$ the first removal state is also strongly renormalized (Fig.~\ref{fig:bandstr_lamd}d). The effective mass of the hole in the first removal state sharply increases above ${\lambda _d} = 0.35$ while for ${\lambda _d} < 0.35$ the effective mass $m_{\|}^*$ along the $\left( {0,0} \right) - \left( {\pi ,\pi } \right)$ direction and $m_ \bot ^*$ along the $\left( {\pi ,0} \right) - \left( {0,\pi } \right)$ direction increase rather weakly with EPI growth (Fig.~\ref{fig:d_effmass}). This sharp increase of the effective mass results from the polaron autolocalization, similar effect has been obtained in Refs.~[\onlinecite{Mishchenko2004,Mishchenko2009}].

In the regime ${\lambda _d} = {\lambda _{pd}} = \lambda $ both the conductivity band and the middle part of the valence band are strongly renormalized by the subbands splitting and spectral weight redistribution (Fig.~\ref{fig:bandstr_lampd}a-d). In this regime we have found the critical coupling value changing in the properties of local single- and two-hole states (see Fig.~\ref{fig:crossover} and Fig.~\ref{fig:fnh2_lpd} above). The sharp increases of the effective mass occurs also at $\lambda  = {\lambda _c}$ (Fig.~\ref{fig:dpd_effmass}), that is the manifestation of the polaron autolocalization.

%===========================================================================
\section{Polaronic spectral function \label{sp_fun}}
%===========================================================================
\begin{figure*}
\center
\includegraphics[width=0.45\linewidth]{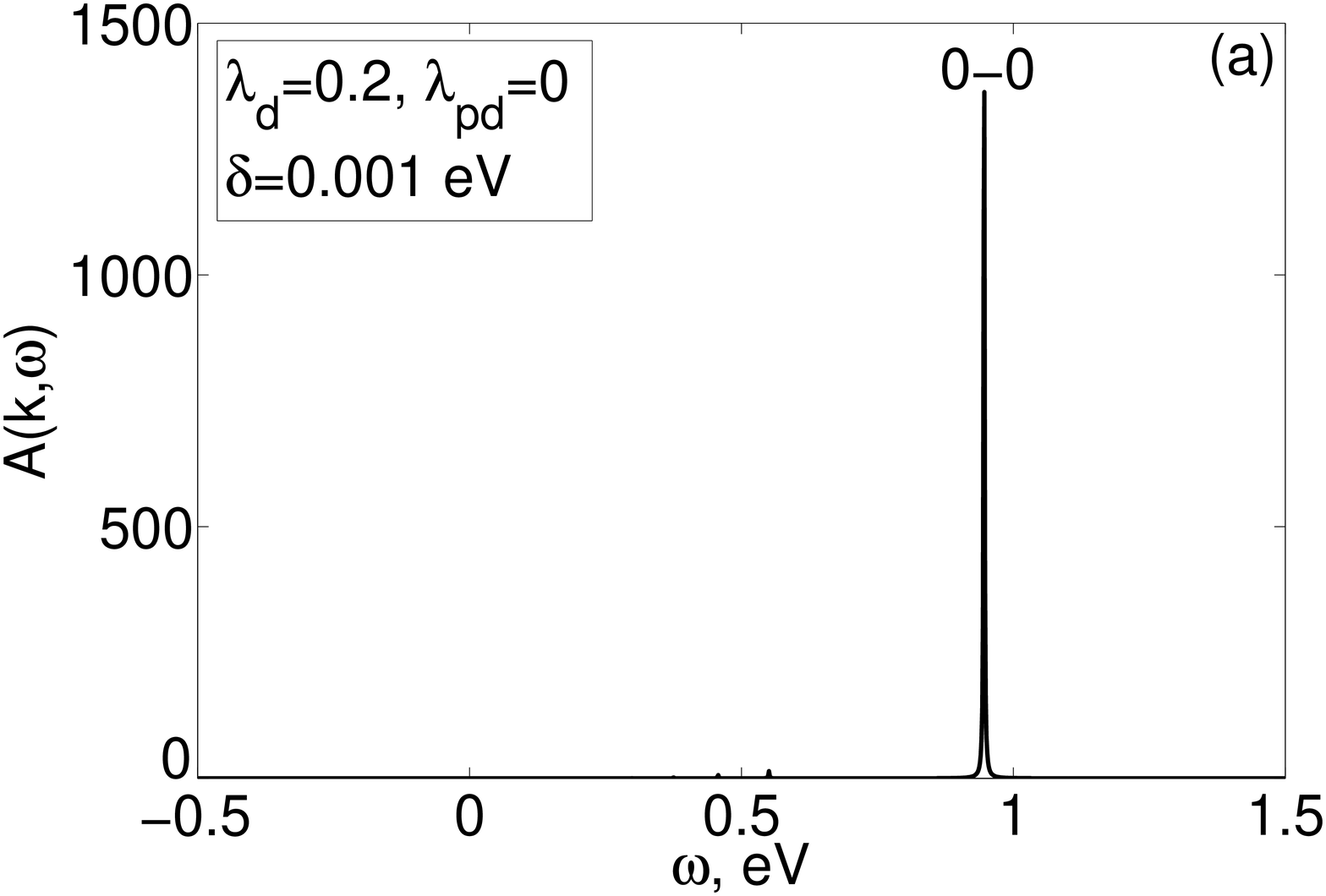}
\includegraphics[width=0.45\linewidth]{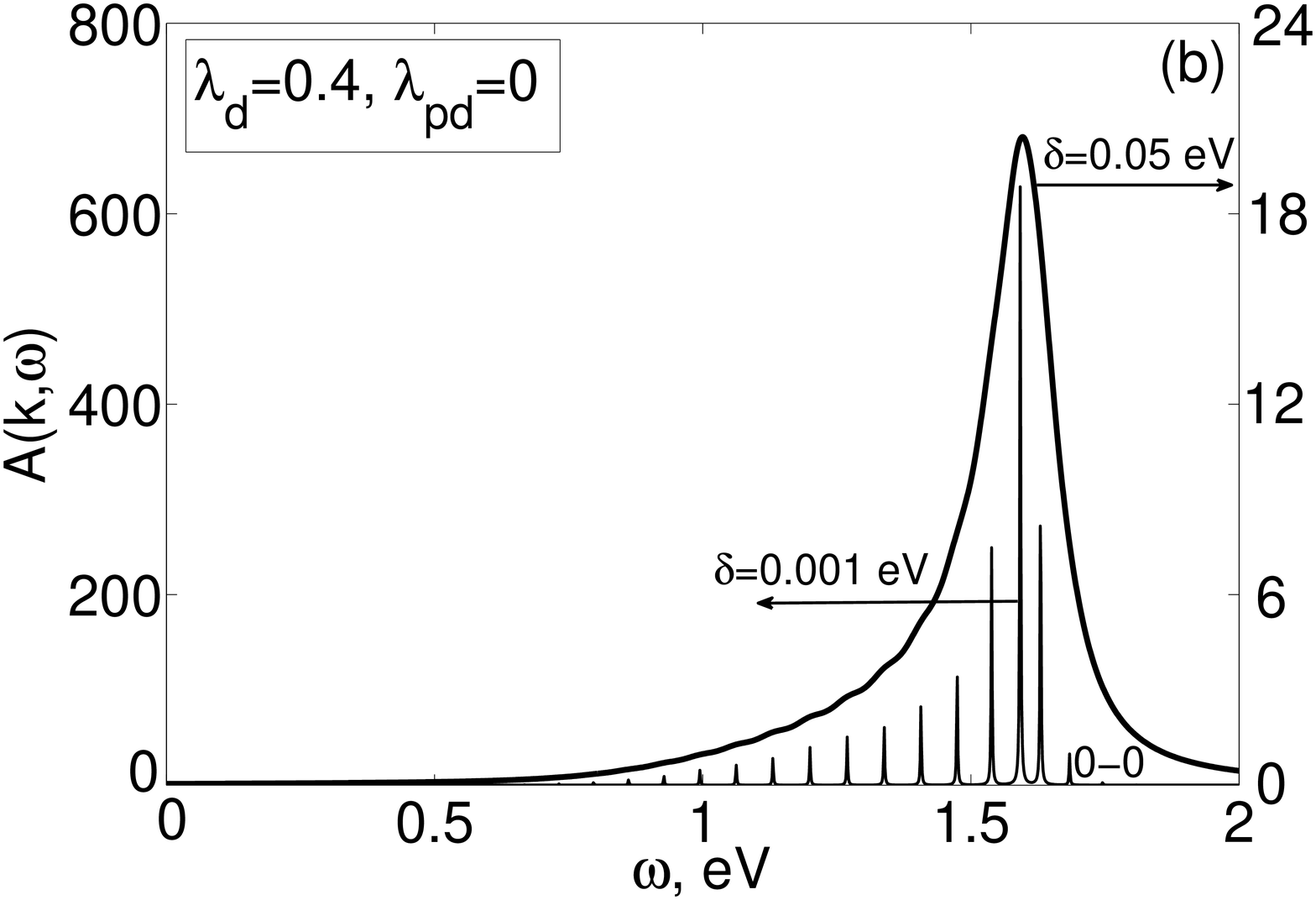}
\includegraphics[width=0.45\linewidth]{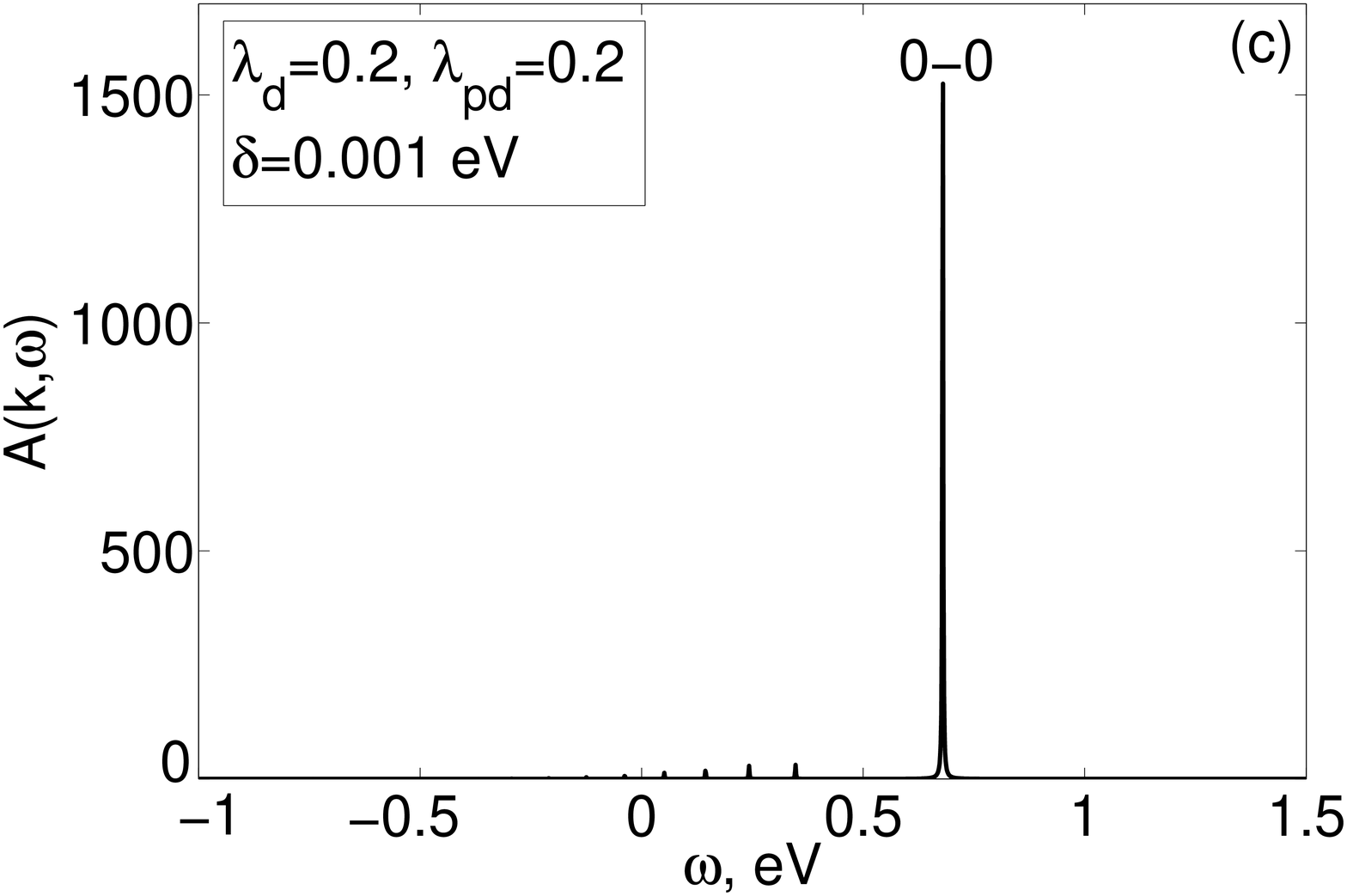}
\includegraphics[width=0.45\linewidth]{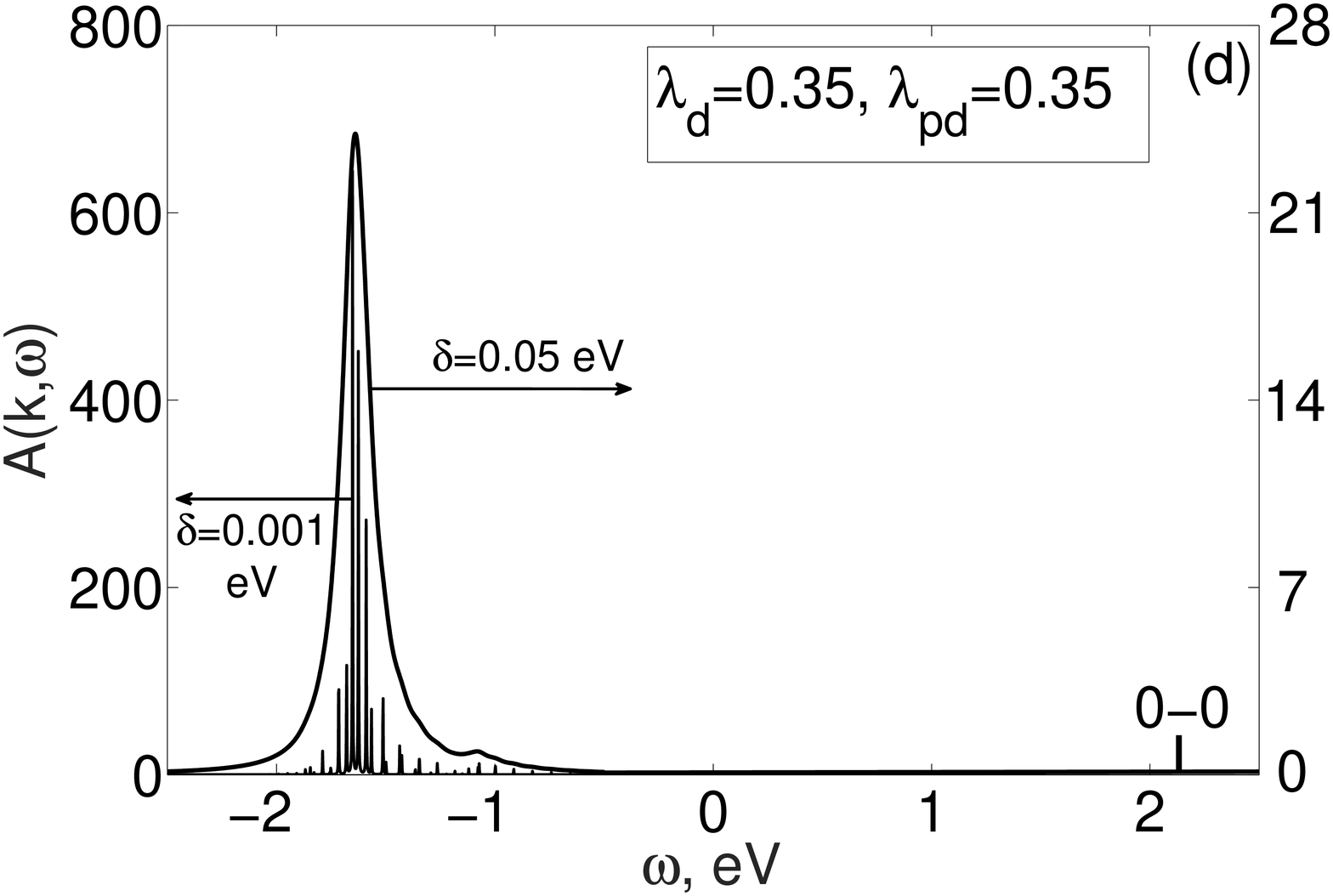}
\caption{\label{fig:Spectral_function} The spectral function of the first removal state at $\left( {\frac{\pi }{2},\frac{\pi }{2}} \right)$ for  ${\lambda _d} = 0.2$ (a), and ${\lambda _d} = 0.4$ with two values of line width $\delta  = 0.001$ eV  and $\delta  = 0.05$ eV (b). The same for ${\lambda _d} = {\lambda _{pd}} = \lambda $ with $\lambda  = 0.2$ (c) and $\lambda  = 0.35$ (d). The multiphonon Franck-Condon resonances are almost absent for delocalized polaron at (a),(c) and appears for localized polaron at (b),(d).}
\end{figure*}

The polaronic spectral function is given by
\begin{equation}
\begin{array}{c}
A\left( {{\bf{k}},\omega} \right) = \sum\limits_\sigma  {{A_\sigma }} \left( {{\bf{k}},\omega} \right) = \\
  - \frac{1}{\pi }\sum\limits_{\sigma \beta mn } {{\gamma _{\sigma \left( \beta \right)}}\left( m \right)\gamma _{ \sigma \left( \beta \right) }^ * \left( n \right){\mathop{\rm Im}\nolimits} {D_{\bf{k}}^{m,n}}\left( {\omega + i\delta } \right)}.
\label{sp_func}
\end{array}
\end{equation}
Splitting of the Hubbard fermion bands to the many hybridized subbands results in the series of narrow peaks for a given wave number. Each peak results from some Franck-Condon resonance and is related to the multiphonon excitation. For ${\lambda _d} < 0.2$ the main peak of the first removal state at the top of the valence band with $ {\bf{k} }= \left( \frac {\pi}{ 2},\frac {\pi}{ 2} \right)$ corresponds to the $0 - 0$ resonance between the $0$-phonon single- and two-hole ground states from Fig.~\ref{fig:hilbert_space}a. The multiphonon contributions to the spectral function are negligibly weak and shifted down in energy (Fig.~\ref{fig:Spectral_function}a). With the diagonal EPI increasing the $0 - 0$ peak decreases while the set of multiphonon peaks appears (Fig.~\ref{fig:Spectral_function}b). For the equal diagonal and off-diagonal EPI situation is qualitatively similar (Fig.~\ref{fig:Spectral_function}c,d). The shift of the multiphonon peak from the $0 - 0$ resonance with almost zero spectral weight is larger in Fig.~\ref{fig:Spectral_function}d.

The effect of the quasiparticle finite lifetime on the spectral function is also shown in Fig.~\ref{fig:Spectral_function}b,d. It is modelling by the different Lorenzian width $\delta$. With increasing $\delta$ we reproduce the formation of one wide peak in the spectral function from the sum of several Franck-Condon resonances. This mechanism of the large linewidth in the ARPES experiments in the undoped cuprate has been discussed in the paper.~\cite{ShenKM2007} The large shift of the spectral intensity below the nominal top of the valence band in the absence of EPI given by $0 - 0$ resonance has been also found in ARPES measurements.~\cite{ShenKM2004}

%===========================================================================
\section{Discussion of the results\label{discussion}}
%===========================================================================
In this paper we have developed the general approach to the electronic structure of Mott-Hubbard insulators with strong electron correlations and strong electron-phonon coupling. The polaronic  version of GTB (or p-GTB) method is a variant of the cluster perturbation theory with exact diagonalization of the intracell part of the Hamiltonian,  redefinition of all local fermionic and bosonic operators as the linear combination of the quasiparticles that are excitations between multielectron and multiphonon initial and final states. The Fermi type excitations are Hubbard polarons. The formally exact generalized Dyson equation has been solved in the conventional for the cluster perturbation theory Hubbard I approximation that results in the polaronic band structure of hybridized Hubbard fermions and local multiphonon Franck-Kondon resonances. The polaronic spectral weight is strongly dependent on the value of the EPI. We have carried out all calculations here within the three band $p-d$ model. It is evident that the p-GTB may be straightforwardly generalized to the multiband realistic model with all copper $d$-orbitals and all oxygen $p$-orbitals, and to the more realistic treatment of the phonon system. The restriction of the Hubbard I approximation is not obligatory for the GTB method, for example the self-consistent calculation of the self-energy and spin correlation functions for the $t-J$ model had been carried out in the X-operators technique in the spin liquid state of the doped cuprates~\cite{KorshOvch2007} and within non-crossing approximation.~\cite{Plakida1999}

The band dispersion and the spectral weight of Hubbard polarons are strongly temperature and doping dependent. The finite temperature generalization of the GTB band structure calculations has been demonstrated recently for the undoped La$_2$CuO$_4$ where the  insulator state is shown to exist both in the antiferromagnetic phase below the Neel temperature and in the paramagnetic phase above the Neel temperature.~\cite{MakOvch2015}
Due to the finite volume of this paper we have restricted ourselves here to only undoped cuprate, the discussion of the effects of doping and finite temperature on the polaronic band structure will be given in a forthcoming paper.

We have shown that in general polaron is characterized by a broad distribution of the phonon numbers in surrounding cloud. Depending on the EPI coupling the maxima in the phonon distribution are given by the $0$-phonon or multiphonon contributions. The former corresponds to the delocalized large polaron while the latter describes the localized small polaron with the crossover from large to small polaron with increasing EPI. Previously similar behavior has been found  by the diagrammatic Monte Carlo method~\cite{CatFilMisNag} within the $t-J$-Holstein model. The new results in our p-GTB approach have been obtained when both diagonal and off-diagonal EPI were considered. We have found their partial compensation in formation the multiphonon cloud when both coupling parameters were equal and the sharp transition from the large to small polaron at the critical value of EPI coupling. This transition is accompanied by the polaron localization and its effective mass divergence. The other our new result in comparison to the $t-J$ Holstein model treatment by preceding authors is the simultaneous transformation of both the valence and conductivity bands under EPI growth and different effect of the diagonal and off-diagonal EPI on these bands.
For the undoped cuprate we have reproduced  the large width of the ARPES line measured for the first removal state related with the loss of the spectral weight by the $0-0$ Franck-Kondon resonance and shift  the spectral weight maximum down in energy to the multiphonon resonances, these features have been found experimentally in the undoped  Sr$_2$CuO$_2$Cl$_2$.~\cite{ShenKM2004,ShenKM2007}

\begin{acknowledgments}
Authors are thankful to Russian Science Foundation (project No. 14-12-00061) for financial support.
\end{acknowledgments}

\appendix

%===========================================================================
\section{Renormalized parameters of the initial Hamiltonian and structural factors}
%===========================================================================
The link between initial parameters and new matrix elements in Eq.~({\ref{Hc_Hcc}}) are given by:
\begin{equation}
\label{link:1}
\varepsilon_b = \varepsilon_p - 2t_{pp}\nu_{00},
\end{equation}
\begin{equation}
\label{link:2}
{U_b} = {U_p}{\Psi _{0000}},
\end{equation}
\begin{equation}
\label{link:3}
{\tilde V_{pd}} = {V_{pd}}{\Phi _{000}}.
\end{equation}
Coefficients $\nu_{00}$, $\Psi _{0000}$, $\Phi _{000}$ are values of structural factors $\nu_{\bf{fg}}$, $\Psi _{\bf{ijkl}}$, $\Phi _{\bf{ijk}}$ for single cluster, $\nu_{00}=0.727$, $\Psi _{0000}=0.2109$, $\Phi _{000}=0.918$. Values of $\Psi _{\bf{ijkl}}$, $\Phi _{\bf{ijk}}$ are strongly decreased with distance increasing.

The structural factors $\mu_{{\bf{fg}}}$, $\nu_{{\bf{fg}}}$, $\rho_{{\bf{fgh}}}^A$ are defined by relations:
\begin{equation}
\label{link:4}
{\mu _{\bf{fg}}} = {1 \mathord{\left/
 {\vphantom {1 N}} \right.
 \kern-\nulldelimiterspace} N}\sum\limits_{\bf{k}} {{\mu _{\bf{k}}}{e^{ - i{\bf{k}}\left( {{\bf{f}} - {\bf{g}}} \right)}}},
\end{equation}
\begin{equation}
\label{link:5}
{\nu _{\bf{fg}}} = {1 \mathord{\left/
 {\vphantom {1 N}} \right.
 \kern-\nulldelimiterspace} N}\sum\limits_{\bf{k}} {{{\left( {{{2\sin \left( {{{{k_x}} \mathord{\left/
 {\vphantom {{{k_x}} 2}} \right.
 \kern-\nulldelimiterspace} 2}} \right)\sin \left( {{{{k_y}} \mathord{\left/
 {\vphantom {{{k_y}} 2}} \right.
 \kern-\nulldelimiterspace} 2}} \right)} \mathord{\left/
 {\vphantom {{2\sin \left( {{{{k_x}} \mathord{\left/
 {\vphantom {{{k_x}} 2}} \right.
 \kern-\nulldelimiterspace} 2}} \right)\sin \left( {{{{k_y}} \mathord{\left/
 {\vphantom {{{k_y}} 2}} \right.
 \kern-\nulldelimiterspace} 2}} \right)} {{\mu _{\bf{k}}}}}} \right.
 \kern-\nulldelimiterspace} {{\mu _{\bf{k}}}}}} \right)}^2}{e^{ - i{\bf{k}}\left( {{\bf{f}} - {\bf{g}}} \right)}}},
\end{equation}
\begin{equation}
\begin{array}{c}
\label{link:6}
\rho _{{\bf{fgh}}}^A = \rho _{{\bf{f}} - {\bf{g}},{\bf{g}} - {\bf{h}}}^A{{ = 1} \mathord{\left/
 {\vphantom {{ = 1} {{N^2}}}} \right.
 \kern-\nulldelimiterspace} {{N^2}}}\sum\limits_{{\bf{kq}}} {{1 \mathord{\left/
 {\vphantom {1 {{\mu _{\bf{k}}}{\mu _{\bf{q}}}}}} \right.
 \kern-\nulldelimiterspace} {{\mu _{\bf{k}}}{\mu _{\bf{q}}}}}}  \times \\
\left[ {\sin \left( {{{{k_x}} \mathord{\left/
 {\vphantom {{{k_x}} 2}} \right.
 \kern-\nulldelimiterspace} 2}} \right)\sin \left( {{{{q_x}} \mathord{\left/
 {\vphantom {{{q_x}} 2}} \right.
 \kern-\nulldelimiterspace} 2}} \right)\cos \left( {{{\left( {{k_x} + {q_x}} \right)} \mathord{\left/
 {\vphantom {{\left( {{k_x} + {q_x}} \right)} 2}} \right.
 \kern-\nulldelimiterspace} 2}} \right) + } \right.\\
\left. {\sin \left( {{{{k_y}} \mathord{\left/
 {\vphantom {{{k_y}} 2}} \right.
 \kern-\nulldelimiterspace} 2}} \right)\sin \left( {{{{q_y}} \mathord{\left/
 {\vphantom {{{q_y}} 2}} \right.
 \kern-\nulldelimiterspace} 2}} \right)\cos \left( {{{\left( {{k_y} + {q_y}} \right)} \mathord{\left/
 {\vphantom {{\left( {{k_y} + {q_y}} \right)} 2}} \right.
 \kern-\nulldelimiterspace} 2}} \right)} \right] \times \\
{e^{ - i{\bf{k}}\left( {{\bf{f}} - {\bf{g}}} \right)}}{e^{ - i{\bf{q}}\left( {{\bf{g}} - {\bf{h}}} \right)}}.
\end{array}
\end{equation}
Their values in depend on distance are given in Table~\ref{mu_nu_po}.
\begin{table*}[htb]
\begin{center}
\begin{tabular}{|*{5}{c|}}
%\begin{tabular}{c@{\hspace{3mm}}c@{\hspace{5mm}}*{12}{r@{\hspace{3mm}}}}
\hline
  $\left( {{f_x}-{g_x},{f_y}-{g_y}} \right)$ & $\left( {{g_x}-{h_x},{g_y}-{h_y}} \right)$ & $\mu_{\bf{fg}}$ & $\nu_{\bf{fg}}$ & $\rho _{\bf{fgh}}^A$ \\
\hline
  $\left( {0,0} \right)$ & $\left( {0,0} \right)$ & 0.958 & 0.727 & -0.459 \\
\hline
  $\left( {1,0} \right)$ & $\left( {0,0} \right)$ & -0.14 & -0.273 & 0.067 \\
\hline
  $\left( {1,1} \right)$ & $\left( {0,0} \right)$ & -0.024 & 0.122 & 0.011 \\
\hline
  $\left( {2,0} \right)$ & $\left( {0,0} \right)$ & -0.014 & -0.064 & 0.007 \\
\hline
  $\left( {0,0} \right)$ & $\left( {1,0} \right)$ & -- & -- & 0.067 \\
\hline
  $\left( {1,0} \right)$ & $\left( {1,0} \right)$ & -- & -- & -0.121 \\
\hline
  $\left( {1,0} \right)$ & $\left( {0,1} \right)$ & -- & -- & 0.023 \\
\hline
  $\left( {1,1} \right)$ & $\left( {1,0} \right)$ & -- & -- & -0.015 \\
\hline
  $\left( {1,0} \right)$ & $\left( {1,1} \right)$ & -- & -- & -0.015 \\
\hline
\end{tabular}
\caption{The values of the coefficients $\mu_{{\bf{fg}}}$, $\nu_{{\bf{fg}}}$, $\rho_{{\bf{fgh}}}$    as a function of the site coordinates $\bf{f}$, $\bf{g}$, $\bf{h}$}
\label{mu_nu_po}
\end{center}
\end{table*}

%===========================================================================
\section{Matrices of intercluster electron-phonon interaction}
%===========================================================================
Matrix of intercluster EPI in the Dyson equation (\ref{DysonEq}) includes two terms:
\begin{equation}
\label{link:7}
\hat M_{\bf{k}}^{EPI} = \hat M_{\left( {pd} \right){\bf{k}}}^{\left( 1 \right)} + \hat M_{\left( {pd} \right){\bf{k}}}^{\left( 2 \right)},
\end{equation}
where $\hat M_{\left( {pd} \right){\bf{k}}}^{\left( 1 \right)}$, $\hat M_{\left( {pd} \right){\bf{k}}}^{\left( 2 \right)}$ - matrices of off-diagonal intercluster EPI.
\begin{equation}
\begin{array}{c}
\label{link:8}
\hat M_{pd\,{\bf{k}}}^{(1)}\left( {uv;nm} \right) = 2{g_{pd}}{\xi _d}\rho _{1{\bf{k}}} \times \\
\sum\limits_p {\gamma _A^ * \left( {pp} \right)\gamma _{{d_x}}^ * \left( {vu} \right){\gamma _b}\left( {nm} \right)\left\langle {{X^{pp}}} \right\rangle }
\end{array}
\end{equation}
with coefficient $\rho _{1{\bf{k}}} = \sum\limits_{\bf{gh}} {\rho _{{\bf{g}}{\bf{h}}}^A{e^{i{\bf{k}}{\bf{h}}}}}$
\begin{equation}
\begin{array}{c}
\label{link:9}
\hat M_{pd\,{\bf{k}}}^{(2)}\left( {uv;nm} \right) = 2{g_{pd}}{\xi _d}\rho _{2{\bf{k}}} \times \\
\sum\limits_p {\gamma _A^ * \left( {pp} \right)\gamma _{{d_x}}^ * \left( {vu} \right){\gamma _b}\left( {nm} \right)\left\langle {{X^{pp}}} \right\rangle }
\end{array}
\end{equation}
with coefficient $\rho _{2{\bf{k}}} = \sum\limits_{\bf{h}} {\rho _{0,{\bf{h}}}^A{e^{i{\bf{k}}{\bf{h}}}}}$

\bibliography{C:/literature_pGTB}

\end{document}